\documentclass[%
reprint,
amsmath,amssymb,
aps,
pra,
floatfix,
]{revtex4-1}

\usepackage{graphicx}% Include figure files
\usepackage{dcolumn}% Align table columns on decimal point
\usepackage{bm}% bold math
\usepackage{color}
\usepackage{soul}
\usepackage{float}

\begin{document}

\title{Ultrafast chirality: the road to efficient chiral measurements}

\author{David Ayuso$^{1,2}$} \email{d.ayuso@imperial.ac.uk}
\author{Andres F. Ordonez$^{1,3}$} \email{andres.ordonez@icfo.eu}
\author{Olga Smirnova$^{1,4}$} \email{olga.smirnova@mbi-berlin.de}

\affiliation{$^1$Max-Born-Institut, 12489 Berlin, Germany}
\affiliation{$^2$Department of Physics, Imperial College London, SW7 2AZ London, United Kingdom}
\affiliation{$^3$ICFO-Institut de Ciencies Fotoniques, The Barcelona Institute of Science and Technology, 08860 Barcelona, Spain}
\affiliation{$^4$Technische Universit\"at Berlin, 10623 Berlin, Germany}

\raggedbottom

\begin{abstract}

Today we are witnessing the electric-dipole revolution in chiral measurements.  Here we reflect on its lessons and outcomes, such as the perspective on chiral measurements using  the complementary principles of ``chiral reagent'' and ``chiral observer'', the hierarchy of scalar, vectorial and tensorial observables, the new properties of the chiro-optical response in the ultrafast and non-linear domains, and the geometrical magnetism associated with the chiral response in photoionization. The electric-dipole revolution is a landmark event. It  has opened routes to extremely efficient enantio-discrimination with a family of new methods. These methods are governed by the same principles but work in vastly different regimes -- from microwaves to optical light; they address all molecular degrees of freedom -- electronic, vibrational and rotational, and use flexible detection schemes, i.e. detecting  photons or electrons, making them applicable to different chiral phases, from gases to liquids to amorphous solids. The electric-dipole revolution has also enabled enantio-sensitive manipulation of chiral molecules with light.  This manipulation includes exciting and controlling ultrafast helical currents in vibronic states of chiral molecules,  enantio-sensitive control of populations in electronic, vibronic and rotational molecular states, and opens the way to efficient enantio-separation and enantio-sensitive trapping  of chiral molecules. The word ``perspective'' has two meanings: ``outlook'' and ``point of view''. In this  perspective article, we have tried to cover both meanings.

\end{abstract}

\maketitle

Chirality is a fundamental concept of geometrical origin ubiquitous in nature, from elementary particles to molecules to macroscopic objects. Chiral molecules are characterised by the spatial arrangement of their nuclei, which makes them non-superimposable on their mirror image.
This property underlies their important function in chemistry and biology: chiral structure facilitates recognition at the molecular level, which is a key component of metabolic reactions in biological organisms.

Most biologically relevant molecules are chiral, and many appear in nature only in one of their two possible forms  \cite{book_Wade_OrganicChemistry}. Whereas the amino acids found in living beings are essentially left-handed, sugars are right-handed. While the origin of biological homochirality, i.e. the single-handedness of key biomolecules, is still debated  \cite{Bonner1991,Cohen1995,Hazen2001,Blackmond2004,Breslow2009,Weissbuch2011,Hein2012,Hein2013,Brewer2014,Jafarpour2015},  its importance could be related to the fact that this geometrically protected quantity -- handedness -- carries a single bit of information \cite{Brewer2014} -- left or right, i.e. ``true'' or ``false''. Then, to enable such information processing in biological systems, the balance between left and right must be broken.  The measure of such imbalance, the enantiomeric excess,
is maximized in homochiral systems, making them ideal for chirality-based information processing.

Since handedness is a key element in molecular recognition, it is both fundamentally interesting and important to learn how it is encoded 
in interactions between chiral molecules and light. 

Many applications use chirality for characterizing molecular structures via linear light-matter interaction. For example, the helical structures of DNA, RNA, and some proteins are routinely characterized by absorption circular dichroism (CD) measurements. The results are typically interpreted on a phenomenological basis, e.g. by comparing to benchmark structures.
The physical understanding of chirality at the molecular level, i.e. at the level of the relevant properties of electronic structure and dynamics, is challenging. Indeed, little 
is known about the dynamics of chirality in terms of concepts such as electronic ring currents \cite{Barth2016,Bredtmann2011} and the fields they may generate inside molecules \cite{Barth2007PRA,Yuan2013PRA}, ultrafast charge migration \cite{Cederbaum1999,Breidbach2003,Kuleff2014,Lunnemann2008,Remacle1999,Remacle2006,Calegari2014,Calegari2016,Nisoli2017,Ayuso2017} or nonlinear electronic responses \cite{Mukamel1995}. Yet, the dynamical response provides a different and independent access to the
physical mechanisms underlying the chiral function. 

While understanding chiral interactions and time-resolving  
chiral electronic or vibronic dynamics are much desired, most of the existing ultrafast  methods are restricted to weak interactions with the magnetic component of the light field (e.g.  \cite{meyer2013recent,ghosh2021broadband,Rouxel2020,Ye2019,Rouxel2019,Rhee2009,Oppermann2019,Cireasa2015,Cho2015,Baykusheva2019}). In the IR-VUV range, such restriction severely limits their efficiency for medium-size molecules or chiral moieties, with useful time-resolved signals often just above the noise
 \cite{Ye2019,Rouxel2019,Rhee2009,Oppermann2019,Cireasa2015}. A recent experiment \cite{Oppermann2019} graphically demonstrates the challenges: the time-resolved CD signal is only a few percent of the static CD and is on the order of the baseline stability of the setup. Developing ultrafast and highly enantio-sensitive approaches, which track electronic/vibronic dynamics without relying on magnetic interactions, is an important challenge.

This challenge has been taken on by the electric-dipole revolution in chiral measurements. It began decades ago \cite{Ritchie1976PRA,Powis2000JCP}, but is steadily picking up the pace now, just as we are writing these words. 
It involves several extremely efficient enantio-sensitive methods that rely on purely electric-dipole interactions and address
electronic \cite{Ritchie1976PRA,Powis2000JCP,Bowering2001PRL,Lischke2004,Turchini2004,Stener2004,Stranges2005,Tommaso2006,Turchini2009,Nahon2015JESRP,Ferre2015,Turchini2017,Fehre2021,Kruger2021,Garcia2013,Lux2012Angewandte,Lehmann2013JCP,Stefan2013JCP,Fanood2014,Fanood2015,Lux2015,Lux2015JPB,Beaulieu2016Faraday,Kastner2016CPC,Miles2017,Kastner2019,Ranecky2022,Comby2016JPCL,Beaulieu2016NJP,Beaulieu2018PXCD,Powis2008,Janssen2014PCCP,Hadidi2018,Demekhin2018PRL,Demekhin2019PRA,Goetz2019PRL,Rozen2019PRX,Giordmaine1965,Fischer2005,Neufeld2019PRX, Ayuso2022OptExp,Ayuso2021Optica},
vibrational  \cite{Garcia2013,Beaulieu2018PXCD}
and rotational \cite{Patterson2013PRL,Patterson2013Nat,Lehmann2018}
degrees of freedom in molecules.
Importantly, the analysis of the already existing methods allows one to 
identify the key common principles underlying all these schemes and 
formulate general requirements for experimental 
setups  \cite{Ordonez2018generalized}, which should help one to 
create new efficient enantio-sensitive approaches, tailored to the 
needs, tasks and means  of a given laboratory.

Here we formulate the key principles underlying efficient and ultrafast 
chiral measurements operating in the electric-dipole approximation.  We also 
describe where we see future challenges and frontiers in developing 
such methods. Throughout the paper we discuss several enantio-sensitive phenomena. Respective discussions can be identified in the text by looking at the corresponding subtitles highlighted in bold (e.g. \textbf{Circular Dichroism}).

Our perspectives include not only an outlook for the future, but also a general view on the basic principles underlying various 
chiral measurements and on how these principles can be used to
develop new measurement approaches and increase their efficiency.
This is why we begin our analysis 
with a tutorial-style introduction into 
the basic principles underlying  
chiral measurements. This analysis is presented in Section 2
and focuses on the interaction of chiral molecules with light (or electromagnetic, EM, fields). 
We introduce the concepts of \emph{chiral reagent} and \emph{chiral observer} as two fundamental principles that allow one to construct and analyse  chiral measurements and establish a generalized perspective on chiral measurements common to  different observation regimes.  

These two principles offer complementary approaches. 
One of them, the principle of chiral reagent, 
involves enantio-sensitive interactions of electromagnetic fields with molecules. Another, the chiral observer, 
does not. We show how both approaches  can be upgraded in their efficiency, i.e. realized within the electric-dipole approximation (Sections 3 $\&$ 4), enabling extremely 
efficient enantio-sensitive signals from gas-phase molecules and 
creating new opportunities for efficient probes of chiral liquids.

Sections 3, 4 and 5 use the concepts introduced in 
Section 2 to develop a framework for understanding and realizing 
chiral measurements. To give a brief idea of other topics covered here, 
we list below some specific questions that we address and highlight 
the most important conclusions stemming from the analysis of these questions. 
\begin{enumerate}
\item What is the structure of generic chiral observables 
(Section 2) and how do 
they map onto the ultrafast response of molecules to light (Section 4)?
We shall see that there exists a hierarchy of chiral observables encompassing scalar, vectorial and tensorial quantities. That is, 
there are infinitely many of them, allowing for significant flexibility in 
constructing chiral measurements, an important asset for future experiments.

\item Do we need light's chirality for chiral measurements (Section 3)? 
We shall point out that light's chirality 
is not necessary: it can be substituted by the chirality of an experimental setup. The flexibility in designing chiral experimental
setups offers additional flexibility for efficient enantio-discrimination. 
 
\item Is there a chirality measure for experimental setups (Section 3)? 
We shall see that such chirality measure does indeed exist.
It can therefore serve as a 
guideline for optimizing the enantio-sensitivity of experimental setups.

\item Can light be chiral already in the electric-dipole approximation? 
What are the chirality measures for light (Section 4)?
We shall see that the chirality measures for matter, light, or experimental setups have 
an identical structure, and that there are arrays of these measures 
with an identical hierarchy, 
from scalars to vectors to tensors of various ranks. In the outcome of 
an experiment, the chirality measure of matter couples to the 
chirality measure of light or to the chirality measure of the 
experimental setup (Section 5).

\item How does geometry affect physics: what is the physical mechanism 
of a chiral response (Section 6)? 
Geometrical properties in real space, such as handedness, are defined by 
electronic and nuclear configurations in molecules or lattice configurations 
in solids. In condensed matter physics it has been recently realized that  the manner in which 
these properties map onto the geometrical properties of 
the Hilbert space is very important. This mapping leads to the concepts of Berry phase   \cite{berry1984quantal} and Berry curvature, topological phases  \cite{kane2005quantum,bernevig2006quantum}, and the general concept of geometrical magnetism in solids, providing a novel framework for understanding electronic response  \cite{resta_macroscopic_1994}. About 20$\%$ of all materials known today have electronic properties dictated by the topology of their electronic wave functions  \cite{Felser}, and quantified by the Berry curvature. Therefore, 
we expect that geometry and topology should also play a 
prominent role in the electronic response of chiral molecules and 
should lead to new enantio-sensitive phenomena  \cite{ordonez_geometric_2021}, especially in the non-linear response. 

Our expectations are based on our recent finding of the similarity 
between the Berry curvature in solids and a 
geometric ‘propensity’ field generated by electronic dynamics 
in chiral molecules  \cite{ordonez_geometric_2021}. This field maps the geometric properties of nuclear arrangements sensed by chiral electron currents onto 
chiral observables in photoionization.  In particular, it controls the 
strength of the enantio-sensitive photoionization current orthogonal to the polarization plane of the driving laser field, known as the photoelectron circular dichroism (PECD) in one photon ionization  \cite{Ritchie1976PRA,Powis2000JCP,Bowering2001PRL,Lischke2004,Turchini2004,Stener2004,Stranges2005,Tommaso2006,Turchini2009,Nahon2015JESRP,Ferre2015,Turchini2017,Fehre2021,Kruger2021,Garcia2013}. This control is formally similar to how the 
Berry curvature controls, e.g., the circular photogalvanic current in 
3D materials lacking inversion symmetry, or the anomalous current in 2D two-band topological materials. We expect that PECD is but a tip of the iceberg: the geometric propensity field (Section 8.3)   provides a guiding principle for uncovering a 
new class of enantio-sensitive vectorial observables  \cite{ordonez_geometric_2021}, akin to the anomalous electron velocity due to the Berry curvature in solids. 

\item How can one develop efficient  schemes to control chiral optical responses (Section 6)?
Section 6 outlines our perspectives on efficient control and manipulation of 
chiral molecules with light, and charts the roads for new and highly enantio-sensitive approaches.

\end{enumerate}
\begin{figure}[h]
\centering
\includegraphics[width=0.5\linewidth, keepaspectratio=true]{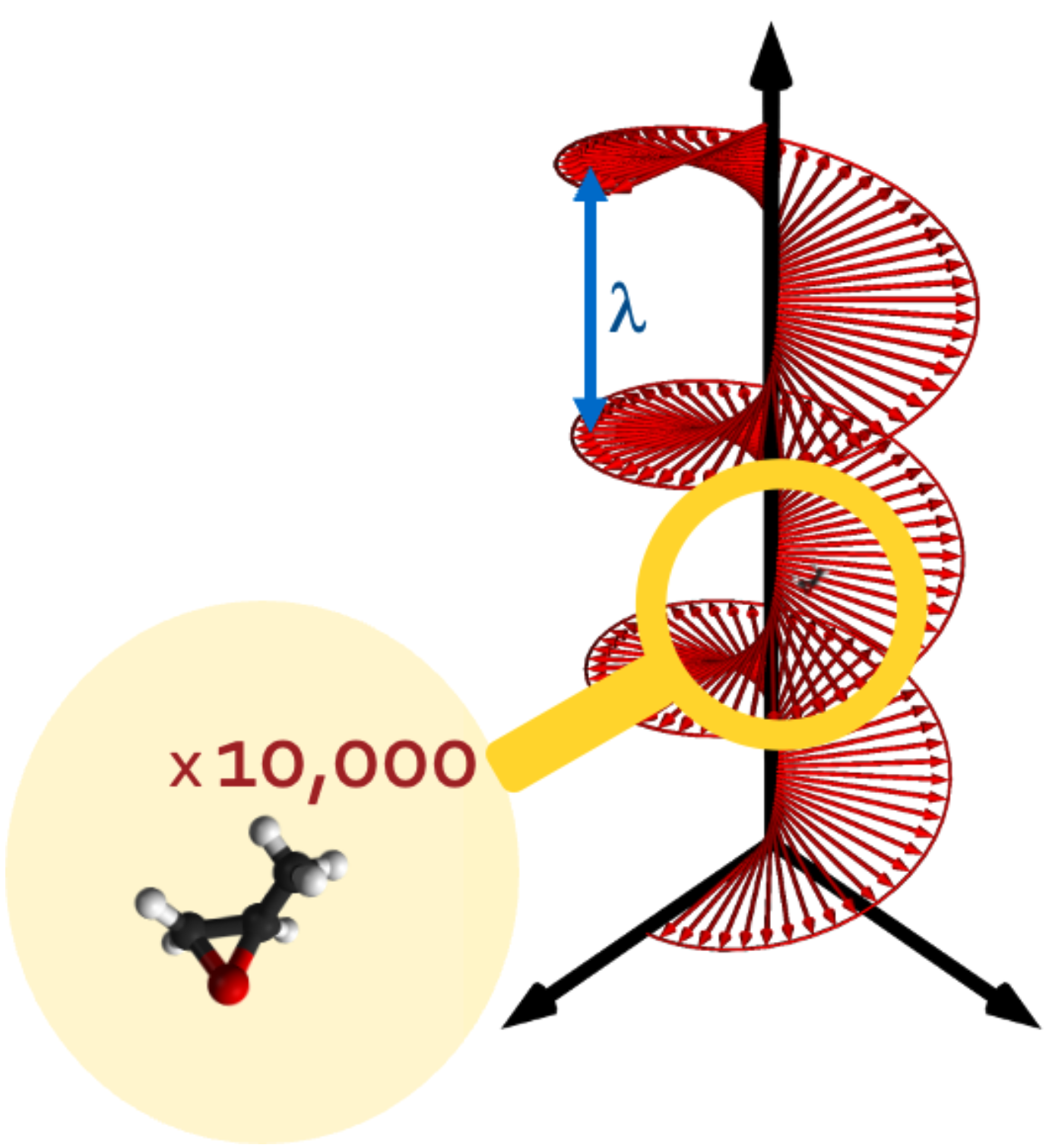}
\caption{Light helix in space. Electric field vector of circularly polarized light makes helix in space. The pitch of the helix is given by light wave-length. For infrared and visible light the helix is orders of magnitude larger than the size of the molecule. It leads to extremely weak enantio-sensitive signals for small molecules, chiral moieties, local chiral centers such as asymmetric carbons.}
\label{Helix_1}
\end{figure} 

\section{Probing chirality: Introducing the concepts of chiral observer and chiral reagent}

How do we find out if  an object is chiral or not, and how do we probe its handedness? 
In the macro-world, we can compare the object with its mirror image.  
This detection principle uses the  concept of \emph{“chiral observer”}; 
the rigorous definition will follow.

Probing chirality can also involve interaction between two chiral objects. One example is a proper handshake, which requires  the two enantiomers -- the two hands -- to have the same handedness. 
This detection principle uses a \emph{“chiral reagent”}, which interacts differently  with left- and right-handed objects. 

In the micro-world, a chiral reagent can  simply be another chiral molecule. The basic reason for the different outcome of a chemical reaction involving chiral molecules is the key difference in their shape. 

Light can also act as a chiral reagent. It is well known that the electric-field vector of a circularly polarized wave draws a basic 
chiral structure in space: a helix (Fig. \ref{Helix_1}). Thus, circularly polarized  light is a chiral photonic reagent, which interacts
differently with left- and right-handed molecules.
The most prominent example being that left- and right-handed molecules absorb different amounts of circularly polarized light of a given handedness, a phenomenon known as photo-absorption circular dichroism (CD).

Absorption CD (e.g.  \cite{Rhee2009,Oppermann2019,Berova2013})  is still the method of choice (e.g.  \cite{Rhee2009,Oppermann2019,Berova2013}) among all-optical methods. Unfortunately, the CD signal is very weak (three to five orders of magnitude smaller than light absorption at the same frequency), as it scales with the ratio of the molecular size to the light’s wavelength (i.e. pitch of the helix).  Optical activity (optical rotation) is a complementary approach based on detecting the rotation of the polarization of a linearly polarized pulse propagating through a chiral medium. It has the same unfavourable scaling  \cite{meyer2013recent,ghosh2021broadband}, especially in the IR-UV range and for small and medium-size molecules. This makes time-resolved measurements very challenging, especially in optically thin media, unless one uses X-ray light to match the pitch of the helix to the molecular size  \cite{Zhang2017,Rouxel2017} or very intense fields  \cite{Cireasa2015,Smirnova_2015JPB,Ayuso2018JPB,Ayuso2018JPB_model,Harada2018,Baykusheva2018PRX} to generate and record high harmonic spectra. Both routes can enhance chiral dichroism from ~0.01$\%$ to a few percent level. 
There are also ingenious field configurations that rely on suppression of the electric-dipole interaction (e.g. \cite{TangCohen, Rosalez-Guzman, forbes_orbital_2021}) to maximise the contribution of "traditional" magnetic-dipole terms, emerging due to the light helix.
Further improvements require one to abandon weak magnetic interactions and rely exclusively on the electric-dipole transitions.

\begin{figure}[h]
\centering
\includegraphics[width=\linewidth, keepaspectratio=true]{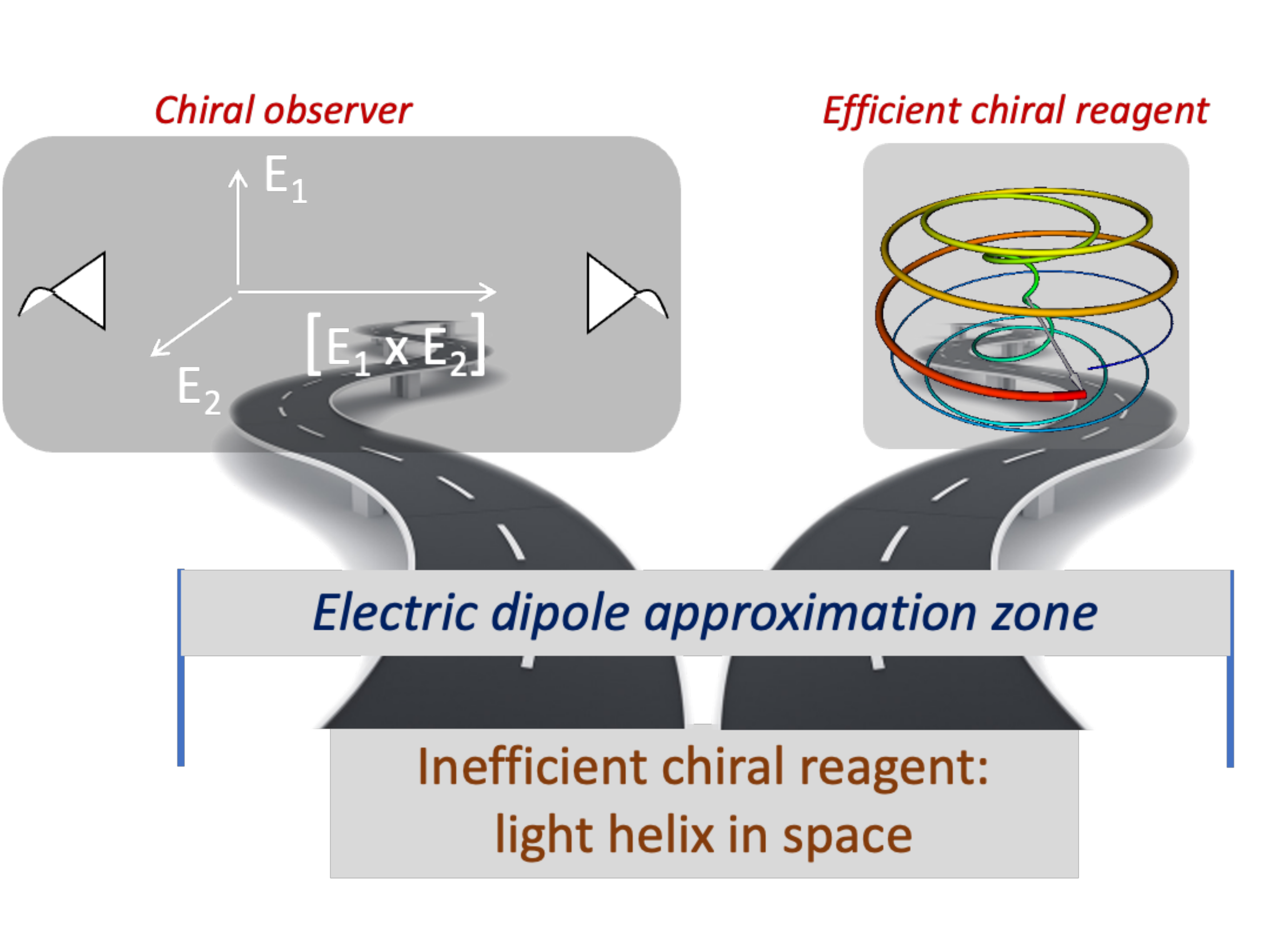}
\caption{Two roads to efficient enantio-discrimination: chiral reagent (approaches that substitute inefficient (non-local) light helix in space by efficient (local) light helix in time) and chiral observer (approaches that do not rely on the chiral properties of light).}
\label{two_roads}
\end{figure} 
There are two roads that one can take (Fig.\ref{two_roads}).  The first option is to develop new approaches that do not rely on the chiral properties of light.
Along this road we will meet a \emph{chiral observer}. The second 
option is to replace the inefficient (non-local)  chiral reagent -- the helix drawn by the light's polarization vector in space -- with an efficient (local) chiral reagent -- a helix, or any other chiral structure, drawn by light's electric vector  in time.  We shall see that these two roads will take us to fundamentally different observables. 
But let us first  recall how one can characterize the handedness of a chiral object.

\subsection{Chiral observables}
\begin{figure}[h]
\centering
\includegraphics[width=\linewidth, keepaspectratio=true]{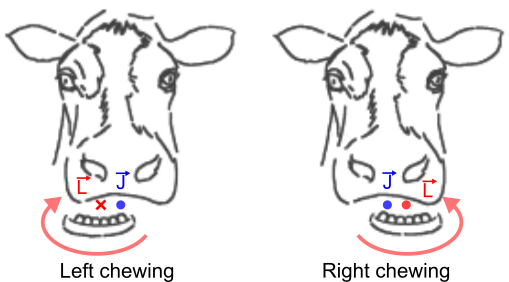}
\caption{Left- and right-chewing (chiral) cows convert in-plane rotation of the jaw, characterized by the pseudo-vector $\vec{L}$, into out-of-plane food motion, characterized by the vector $\vec{J}$.
Reprinted from  \cite{Bungo1999}, with permission from Elsevier.
}
\label{fig_cow}
\end{figure}

Let us consider left-chewing and right-chewing cows  \cite{Jordan1927,Bungo1999} as prototypical carriers of  chiral dynamics\footnote{ We gratefully acknowledge  Prof. D. Herschbach,  who brought to our attention the idea of chiral cows  and its graphical illustration by E. Heller during one of his visits to  Fritz-Haber Institute.}, and define a number that characterizes its handedness. 
While the jaws of a cow make a rotating motion, the food moves in the 
direction orthogonal to the plane of this rotation, see Fig. \ref{fig_cow}.
The chiral dynamics emerging in this process is associated with two directions.  The first is the direction in which food goes, which is a vector $\vec{J}$. The second is the pseudo-vector of the angular momentum $\vec{L}$ associated with the rotation of the jaw. 
An example of a chiral observable, $h$, is a number incorporating these two key directions: a scalar product of $\vec{J}$ and $\vec{L}$, 
\begin{eqnarray}
&&h=\vec{J}\cdot \vec{L}\ \ .
\label{eq:Cows}
\end{eqnarray}
As $\vec{L}$ is a pseudo-vector and $\vec{J}$ is a vector, $h$ is a pseudoscalar. Since a  pseudovector can be generated by taking  a vector product of two vectors, e.g. $\vec{L}=[\vec{r}\times\vec{p}]$, the chiral observable can also be given by a triple product of three vectors.
Pseudo-scalars change sign upon mirror reflection and are, therefore, ideally suited to discriminate opposite enantiomers.

An important conclusion from the above example is that a chiral system can convert in-plane rotation into a linear motion orthogonal to this plane. We shall see that this
can lead to helical electron currents excited in molecules by circualrly polarized electric field. 
It is this coupling that ``merges'' the pseudovector  $\vec{L}$ with the vector $\vec{J}$ into a chiral molecular pseudoscalar $h$. Often the pseudovector component $\vec{L}$ 
is responsible for encoding the in-plane rotation induced e.g. by circuarly polarized electric field, while the vectorial component $\vec{J}$ 
can be used to read out the enantio-sensitive signal, provided that the chiral reference frame (chiral observer) is defined in the experimental set-up. 

\subsubsection{Hierarchy of scalar,  vectorial and tensorial observables}
Experiments always measure ``clicks'', which are scalars. Then, how are pseudoscalars encoded in experimental observables? 
Let us consider the two oldest and most common optical techniques of chiral discrimination, photo-absorption circular dichroism and optical rotation, from this perspective. 

\begin{figure}[h]
\centering
\includegraphics[width=\linewidth, keepaspectratio=true]{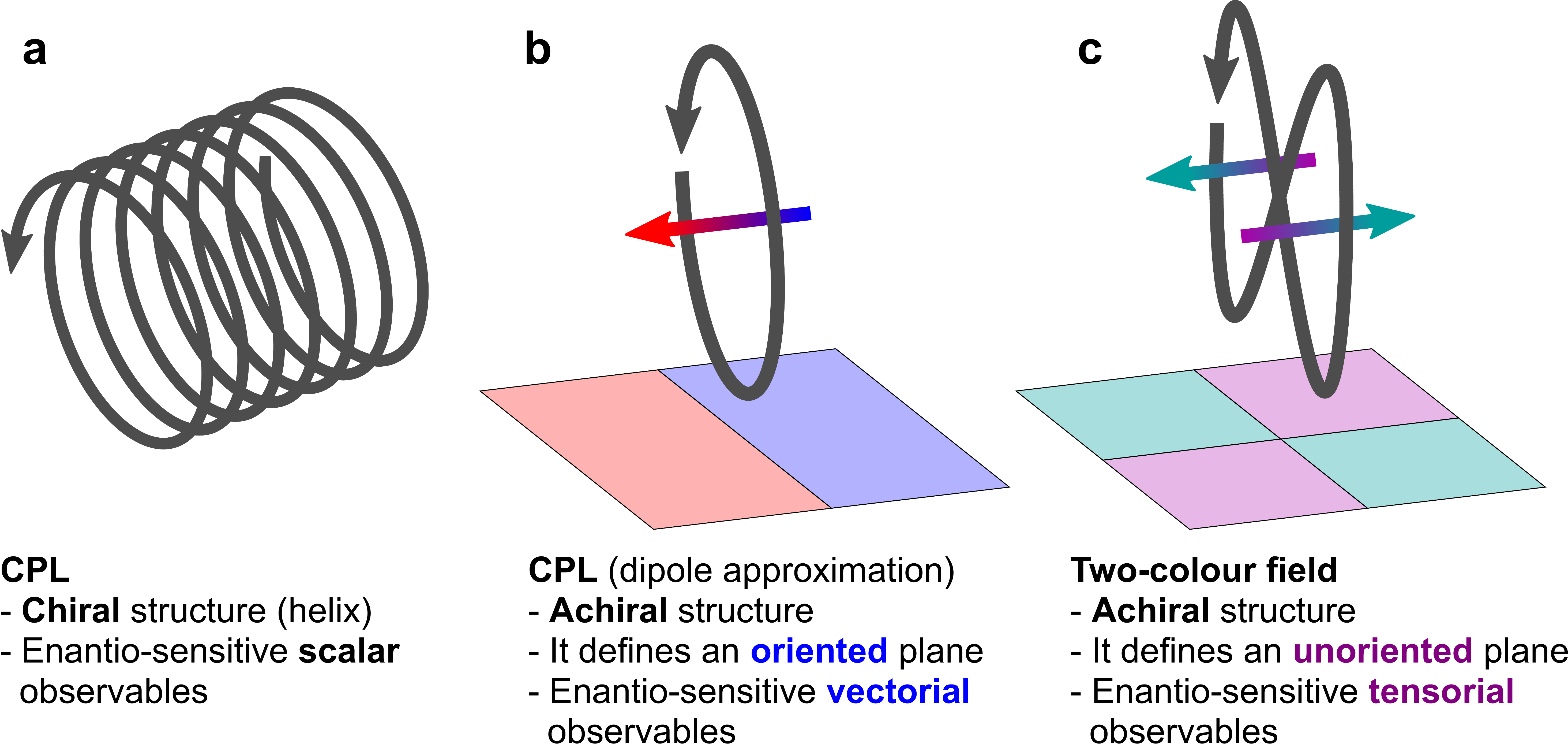}
\caption{\textbf{}
\textbf{a,} Circularly polarized light (CPL) defines a (chiral) helix in space and thus it can lead to enantio-sensitive \emph{scalar} observables.
\textbf{b,} In the electric-dipole approximation, where the helical structure of CPL is neglected, the in-plane rotation of the electric-field vector defines a pseudo-vector, which can couple to molecular pseudoscalars  and lead to enantio-sensitive \emph{vectorial} observables.
\textbf{c,} The Lissajous figure of a two-colour field with orthogonally polarized $\omega$ and $2\omega$ frequency components does not define an oriented plane, and therefore it does not provide a pseudo-vector. It can lead to enantio-sensitive \emph{tensorial} observables.}
\label{fig_scheme}
\end{figure}

\textbf{Absorption Circular Dichroism.} As we have already pointed out, circularly polarized light  can serve as a chiral reagent (Fig. \ref{fig_scheme}a). 
Such a ``reaction'' involves absorption of light, and it is different in the two mirror-reflected versions of a chiral molecule.
Note that the intensity of absorption is a scalar. However, this scalar is, in fact, a product of two pseudoscalars: one associated with the molecule and another associated with the light field. Light's pseudoscalar is given by the optical chirality  \cite{Tang2010}, which is proportional to light's helicity in the case of circularly polarized fields. 
The molecular pseudoscalar is given by the scalar product of the electric-dipole and magnetic-dipole vectors. 
Thus, we need a chiral reagent, in this case the chiral light,  
to ``hide'' a pseudo-scalar inside a scalar. Chiral reagents
allow us to measure  \emph{scalar} enantio-sensitive observables $\blacksquare$.

How does the situation change if we have a setup that allows us to measure \emph{vectorial} observables, such as the polarization of light transmitted through a medium of randomly oriented chiral molecules? The canonical example of such a setup is the 
detection of  chirality via optical rotation. Optical rotation, observed by Biot in 1815, was the first experiment that revealed molecular chirality, but its deep lessons have become especially pertinent now in the light of the electric-dipole revolution. Namely, optical rotation does not use chiral light -- it uses linearly polarized light and detects the rotation of its polarization. 

\textbf{Optical Activity (Optical Rotation).}
When a plane wave of linearly polarized light passes through a chiral medium, its plane of polarization is rotated by an angle that has the same magnitude but opposite signs in opposite enantiomers. Clearly, the characterization of this effect requires (i) specifying light's propagation direction $\hat{k}$ and (ii) the ability to distinguish positive from negative rotations around the polar vector $\hat{k}$. Since rotations are characterized by pseudovectors, this ability requires the definition of a unitary rotation pseudovector $\hat{\theta}$ such that $\hat{\theta}\cdot\hat{k}\neq0$. Introducing polarizer, rotating in the plane orthogonal to the light propagation vector defines such pseudovector, because the direction of polarizer  rotation defines an oriented plane.

When taken together, the vector $\hat{k}$ associated with light and the pseudovector $\hat{\theta}$ associated with 
rotation of polarizer in the laboratory reference frame form a chiral setup. The handedness of this setup is given by $\hat{\theta}\cdot\hat{k}$. 

This handedness of the setup emerges naturally within a formal description of the phenomenon. Indeed, such a description yields an angle of rotation $\vec{\delta}=\delta\hat{k}$, where $\delta\propto\sum_{b}\left(\vec{d}_{a,b}\cdot\vec{m}_{b,a}\right)\big/\left(\omega_{ba}^{2}-\omega^{2}\right)$ is a molecular pseudoscalar with opposite signs for the opposite enantiomers, $a$ and $b$ denote the ground and excited states, respectively, the sum is taken over all excited states, 
$\vec{d}_{a,b}$ and $\vec{m}_{b,a}$ are the transition matrix elements for the electric and magnetic dipole operators $\vec{d}$ and $\vec{m}$, respectively, $\omega$ is the light  frequency and $\omega_{ab}\equiv \omega_{b}-\omega_{a}$. In writing $\vec{\delta}$ vectorially, we have explicitly indicated that the rotation of the plane of polarization (measured by the polarizer $\hat{\theta}$) is defined with respect to $\hat{k}$. 

As for any vectorial measurement, what the experiment reveals are its components with respect to some reference frame. In our case, the relevant unitary (pseudo) vector in our reference frame is $\hat{\theta}$ so that the measurement yields the scalar quantity $\delta_{\theta}=\hat{\theta}\cdot\vec{\delta}\propto\delta(\hat{\theta}\cdot\hat{k})$. That is, the rotation angle $\delta_{\theta}$ measured in the chosen reference frame is the product of two pseudoscalars, one associated with the molecule ($\delta$) and the other associated with the \emph{setup} ($\hat{\theta}\cdot\hat{k}$). If either the handedness of the molecules or of the setup is reversed, $\delta_{\theta}$ changes sign.

Neither the linearly polarized light, nor the polarizer are chiral on their own. Separately, they are not suited to detect chirality, but together they form the chiral setup, or the chiral reference  frame of the observer. Thus, vectorial observables allow one to substitute a \emph{chiral reagent} by a \emph{chiral observer}, i.e. a chiral setup formed by any combination of fields and detectors.$\blacksquare$

One does not have to stop at vectorial observables. 
A chiral observer can also detect tensorial observables of rank higher than one. The hierarchy continues  \textit{ad infinitum} as the tensor rank increases.
Let us consider an example of a chiral setup, which allows one 
to measure vectorial chiral observables using circularly polarized light, 
already in the electric-dipole approximation,  
and then discuss the modifications of the experimental
setup, which allows one to measure tensorial observables. 

Importantly, circularly polarized light in the electric-dipole approximation is not chiral: the spatial helix drawn by the rotating electric-field vector as the light propagates in space (Fig. \ref{fig_scheme}a) is lost
in the electric-dipole approximation. Geometrically, in this 
approximation, the rotation of the electric field vector only 
defines an oriented plane (Fig. \ref{fig_scheme}b). The light supplies a
pseudovector, which defines plane's orientation and is
the vector product of the two orthogonal components of 
the light's polarization. 
To compose a chiral observer, one needs to complement this 
pseudovector with a vector 
orthogonal to the polarization plane.
This vector can be supplied by the detector axis.

When the light field can provide a pseudovector, such as 
the vector product of its two orthogonal polarization components,   the enantio-sensitive and dichroic signal will be a vector collinear with 
the light pseudovector.
But what if the the light field does not define an oriented plane, and therefore it cannot provide a pseudovector?  

A two-color field with orthogonal $\omega$ and $2\omega$ polarizations is one example  of such a field (Fig. \ref{fig_scheme}c). 
Its Lissajous figure changes the direction of rotation twice per laser period. Such a field does not define an oriented plane and does not provide a pseudovector. Yet, it  can still be employed for  enantio-discrimination  \cite{Demekhin2018PRL,Goetz2019PRL,Rozen2019PRX} using a chiral observer. In this case, the chiral observer has to  supply not one, but two detector axes. The first one should be along the direction orthogonal to the polarization plane;  the second one should break the symmetry between the two counter-rotating parts of the figure eight (see Fig. \ref{fig_scheme}c), i.e. should be directed along the polarization of the fundamental field $\vec{E}_{\omega}$.
Thus, the chiral observer can employ two axes defined by the detector.
These two axes define a \emph{quadrupolar} detector, which allows one to correlate measurements in two detection directions. The quadrupolar detector identifies positive or negative correlations between the observables measured along two orthogonal axes (see Section 2.3).   
Such correlation of different detection directions results in  
tensorial chiral observables  \cite{ordonez_molecular_2020}.

The concept of chiral observer embodies a powerful principle of detecting chirality. It allows one to detect chirality using a collection of two 
(or more) non-chiral objects, which are able to collectively form a single chiral object -- a chiral reference frame,  which is used to detect a vectorial (or tensorial) observable. Of course, none of these objects interacts with a chiral molecule in an enantio-sensitive way, as a chiral reagent does. 

The power of the chiral observer comes from the freedom to compose this observer from any vectors available in the experiment, and 
the general framework for designing such measurements.  The measurement no longer has to rely on light's magnetic field to probe chiral molecules, eschewing the need to focus on weak non-electric-dipole effects
and taking maximum advantage of available tools and 
experimental specifics. 

\subsection{Two electric-dipole revolutions in chiral measurements}

Revolutions often come in sequence.  The \emph{first} electric-dipole revolution in chiral measurements has ``revolutionized'' the observer: non-chiral light, interacting with matter in the electric-dipole approximation, in combination with the detection system, has become a very efficient tool for chiral discrimination \cite{Ritchie1976PRA,Powis2000JCP,Bowering2001PRL,Lischke2004,Turchini2004,Stener2004,Stranges2005,Tommaso2006,Turchini2009,Nahon2015JESRP,Ferre2015,Turchini2017,Fehre2021,Kruger2021,Garcia2013,Lux2012Angewandte,Lehmann2013JCP,Stefan2013JACP,Fanood2014,Fanood2015,Lux2015,Lux2015JPB,Beaulieu2016Faraday,Kastner2016CPC,Miles2017,Kastner2019,Ranecky2022,Comby2016JPCL,Beaulieu2016NJP,Beaulieu2018PXCD,Powis2008,Janssen2014PCCP,Hadidi2018,Demekhin2018PRL,Demekhin2019PRA,Goetz2019PRL,Rozen2019PRX,Giordmaine1965,Fischer2005,Neufeld2019PRX, Ayuso2022OptExp,Ayuso2021Optica,Patterson2013PRL,Patterson2013Nat,Lehmann2018}. Since the concept of chiral observer is completely general, it can be applied in a variety of experiments. Such experiments can detect photo-electrons  \cite{Ritchie1976PRA,Powis2000JCP,Bowering2001PRL,Lischke2004,Turchini2004,Stener2004,Stranges2005,Tommaso2006,Turchini2009,Nahon2015JESRP,Ferre2015,Turchini2017,Fehre2021,Kruger2021,Garcia2013,Lux2012Angewandte,Lehmann2013JCP,Stefan2013JCP,Fanood2014,Fanood2015,Lux2015,Lux2015JPB,Beaulieu2016Faraday,Kastner2016CPC,Miles2017,Kastner2019,Ranecky2022,Comby2016JPCL,Beaulieu2016NJP,Beaulieu2018PXCD,Powis2008,Janssen2014PCCP,Hadidi2018,Demekhin2018PRL,Demekhin2019PRA,Goetz2019PRL,Rozen2019PRX}, or molecular fragments  \cite{Pitzer2013Science,Pitzer2017JPB}, or photons, such as optical light  \cite{Giordmaine1965,Fischer2005}, microwave emission  \cite{Patterson2013PRL,Patterson2013Nat,Lehmann2018}, etc. 
Below we shall briefly outline  several ``chiral observer''
methods developed so far, from photoionization to microwave detection to non-linear optics. 

Yet, there is one thing that a chiral observer cannot do -- it cannot detect enantio-sensitive scalar observables. This is where the second electric dipole revolution comes into play. It aims to revolutionize the chiral reagent, i.e. to create and use light \cite{Ayuso2019NatPhot,Ayuso2021NatComm,Ayuso2021ArXiv,Mayer2021ArXiv,katsoulis_momentum_2021,Khokhlova2021ArXiv} or a combination of electric fields  \cite{Gerbasi2004JCP,Yachmenev2019PRL,Eibenberger2017PRL} that is 
chiral already in the electric dipole approximation. This means that the electric field vector of such light, at any given point in space, should  draw a  chiral Lissajous figure during its oscillation period. 
We shall refer to it as ``synthetic chiral light''  \cite{Ayuso2019NatPhot} to distinguish it from  
standard chiral light, i.e. standard
light helix in space.

Since synthetic chiral light (or chiral combination of electric fields including static or microwave electric fields) possesses its own pseudoscalar, it can provide access to scalar observables such as enantio-sensitive populations of molecular electronic, vibronic and rotational states. 
Enantio-sensitive control of populations opens a way to extremely efficient manipulation of chiral molecules such as e.g. enantio-separation,  enantio-sensitive trapping, and possibly even to enantio-sensitive cooling if (or when)
the huge challenges in cooling large molecules are resolved.

The pseudoscalar of synthetic chiral light does not include its magnetic-field component: it is constructed from at least three non-coplanar electric-field components. Thus, the electric-field vector of such light needs to be 3D, a property that may arise in non-collinear configurations \cite{Neufeld2019PRX,Ayuso2022OptExp,Ayuso2019NatPhot,Ayuso2021NatComm,Ayuso2021ArXiv,Mayer2021ArXiv,katsoulis_momentum_2021}, in tightly focused laser beams \cite{Ayuso2021Optica,Khokhlova2021ArXiv} or inside nano-photonic structures \cite{Lodahl2017}, thanks to the emergence of longitudinal components \cite{Bliokh2015}.
What's more, it has to be multi-color, since the three electric field vectors have to be distinguishable. It also means that such light can sense the chirality of matter only via non-linear interactions, as it has to encode the three field
components into the light-induced transition. 

Importantly, the pseudoscalar -- the chirality measure $h$ of synthetic chiral light -- is local. It is defined in the 
electric-dipole approximation and therefore at every point in space: $h=h(\vec{r})$. To control total enantio-sensitive absorption in the entire sample, synthetic chiral light should maintain its handedness globally, across the whole interaction region. That is, its pseudoscalar should maintain its sign from one spatial point to another,  
or at the very least be non-zero on average in space. This can be achieved in different ways, one of them is discussed in Section 4. 

If the light field is not globally chiral, i.e. its pseudoscalar $h=h(\vec{r})$ is zero on average, $\int h(\vec{r})d\vec{r}=0$, higher multipoles of light's handedness will start to play a key role.
For example, we shall see that the presence of the chirality dipole  $\int \vec{r}h(\vec{r})d\vec{r}\ne0$, 
which characterizes the spatial distribution of local handedness in a locally chiral field with $\int h(\vec{r})d\vec{r}=0$, would lead  to enantio-sensitive emission direction of light generated from the molecular sample via non-linear processes such as even harmonic generation \cite{Ayuso2021NatComm}.  

Looking broadly, sculpting multi-color light beams in three dimensions and applying them to chiral molecules appears as a natural next step for the rapidly developing field of generating and using vector beams. This step is compelled by the rich opportunities arising from the interaction of such synthetic chiral light with chiral matter.

\section{The first electric-dipole revolution in chiral measurements: Efficient Chiral Observer} \label{sec:chiral_observer}

The ``chiral observer'' is a concept of chiral measurement in which the detection relies not on the interaction of two chiral objects, but rather on the interaction of a chiral object with two or more achiral objects arranged in such a way that together they form a chiral experimental setup, i.e. the experimental setup becomes a chiral structure. A chiral observer can detect enantio-sensitive vectorial  \cite{Ordonez2018generalized}  and tensorial  \cite{ordonez_disentangling_2020} observables. In the simplest case of vectorial obsevables, the chiral setup  provides a  reference frame, right or left, defined by three non-coplanar vectors. The chiral measurement has to be resolved in this reference frame. Tensorial chiral observables couple to chiral tensorial detectors, correlating two or more detection directions in an enantio-sensitive way.  Enantio-sensitive scalar observables, such as the total intensity of a signal, cannot be produced or measured  in such a setup, since the interaction of two chiral objects is not involved.

 A chiral observer does not require chiral light, and therefore the chiral optical enantio-sensitive response can be induced via strong electric-dipole  interactions, leading to highly efficient enantio-sensitive signals. This is the essence of the first electric-dipole revolution: strong  enantio-sensitive response without chiral light. It has brought a family of new methods benefiting from highly-efficient enentio-sensitive signals based on detecting electrons or photons: PECD in one-photon  \cite{Ritchie1976PRA,Powis2000JCP,Bowering2001PRL,Lischke2004,Turchini2004,Stener2004,Stranges2005,Tommaso2006,Turchini2009,Nahon2015JESRP,Ferre2015,Turchini2017,Fehre2021,Kruger2021,Garcia2013} and multi-photon regimes  \cite{Lux2012Angewandte,Lehmann2013JCP,Stefan2013JCP,Fanood2014,Fanood2015,Lux2015,Lux2015JPB,Beaulieu2016Faraday,Kastner2016CPC,Miles2017,Kastner2019,Ranecky2022,Comby2016JPCL,Beaulieu2016NJP}, PXECD and  PXCD  \cite{Beaulieu2018PXCD}, three wave mixing  \cite{fischer_chiral_2002,fischer_isotropic_2001,fischer_new_2003, Fischer2005,belkin_sum-frequency_2000} and enantio-sensitive microwave spectroscopy  \cite{Patterson2013Nat,Eibenberger2017PRL}. The relevant vectorial observables are the induced polarization or photoelectron current. Tensorial observables related to multipolar bound polarizations \cite{ordonez_inducing_2021} or multipolar photoelectron currents have been predicted in the two-photon regime  \cite{Demekhin2018PRL,Demekhin2019PRA} and measured in the multiphoton regime  \cite{Rozen2019PRX}. 
 
 The concept of chiral observer provides a set of rules to find new vectorial or tensorial enantio-sensitive observables, and to design the experimental setup required for their observation.
These rules may also allow us to adapt the setup to the specifics of the experimental tools available in each laboratory.
Let us formulate these simple rules.

\smallskip
\textbf{Rule 1:}
\textsf{The enantio-sensitive vectorial signal excited by the light field in a randomly oriented ensemble of chiral molecules and detected by a chiral observer has the following general form:}
\begin{equation}
    \langle\vec{v}\rangle =g \vec{L},
    \label{eq:v=gL}
\end{equation}
\textsf{where $g$ is a molecular pseudoscalar and $\vec{L}$ is a field pseudovector.}
\smallskip

The lowest-order field pseudovector that can be constructed without the help of the magnetic component of the light field is the vector product of the 
two orthogonal, phase-delayed electric-field components. 
Such a vector product arises naturally in elliptically polarized fields, maximizing in circularly polarized fields. The respective molecular pseudoscalar is a triple product of three vectors: the first two are the non-collinear dipole moments responsible for the coupling with each of the two field components present in the light pseudoscalar. The third is a vector $\vec{v}$ representing the desired vectorial observable.

The two types of  vectorial observables  outlined so far require observation of light generated due to the induced dipole $\vec{d_{if}}$, 
$\vec{v}=\vec{d_{if}}$ (e.g. PXCD, see below), or ejected photoelectrons with momentum $\vec{k}$, $\vec{v}=\vec{k}$ 
(e.g. PECD, see below). Thus, the molecular pseudoscalar $g$ includes the triple product of dipoles $[\vec{d_{ik}}\times\vec{d_{ki}}]\cdot\vec{d_{if}}$ for detecting light and a similar construct in the case of detecting photoelectrons  $[\vec{d_{ik}}\times\vec{d_{*ki}}]\cdot\vec{k}$. The two pseudoscalars differ by a single vector - the one characterizing the type of observation.
 
To illustrate \textbf{Rule 1}, consider the phenomenon that heralded the electric-dipole revolution, photoelectron circular dichroism (PECD) in one-photon ionization by circularly polarized light.
  
\textbf{Photoelectron circular dichroism, PECD,}
was predicted by Ritchie \cite{Ritchie1976PRA}, Cherepkov  \cite{cherepkov_angular_1972,cherepkov_angular_1981} and Powis  \cite{Powis2000JCP} and first detected by B\"owering et al  \cite{Bowering2001PRL}. It has now 
been extended to the multiphoton  \cite{Lux2012Angewandte,Lehmann2013JCP} and strong-field ionization regimes  \cite{Beaulieu2016NJP} and is being adopted for industrial applications  \cite{powis2017direct}.

In PECD, the circularly polarized field
 \begin{equation} 
 \vec{E}(t)=\vec{E}_{\omega}e^{-i\omega t}+\mathrm{c.c.} 
 \end{equation}
 ionizes an isotropic sample of randomly oriented chiral molecules.
The resulting photoelectron angular distribution (averaged over all molecular orientations) is
found to be asymmetric with respect to the polarization plane of the light. This 
so-called forward-backward asymmetry amounts to 
the generation of a net photoelectron current perpendicular to the polarization plane, which 
displays both enantio-sensitivity and circular dichroism. 

The PECD current can be written as \cite{Ordonez2018generalized} 
 \begin{equation}
 \vec{j}=g \vec{L}, 
 \end{equation}
 where
 \begin{equation}
 \label{PECD_mol_pseudoscalar}
 g\equiv\frac{1}{6}\int \mathrm{d}\Omega_k (\vec{d}_{\vec{k},i}^{*}\times\vec{d}_{\vec{k},i})\cdot\vec{k}, 
 \end{equation}
 $\int \mathrm{d}\Omega_k$ indicates integration over all directions of the photoelectron momentum $\vec{k}$, $\vec{d}_{\vec{k},i}\equiv\langle\vec{k}|\vec{d}|i\rangle$ is the transition dipole matrix element between the ground state $|i\rangle$ and the continuum state $|\vec{k}\rangle$, and 
 \begin{equation}
 \vec{L}\equiv \vec{E}^*_{\omega}\times\vec{E}_{\omega}.
 \end{equation}
Using the expression linking photoionization dipoles for left and right enantiomers  $\vec{d}_{\vec{k},i}^{\mathrm{left}}=-\vec{d}_{-\vec{k},i}^{\mathrm{right}}$, one can show that $g$ has opposite values for opposite enantiomers \cite{Ordonez2018generalized}, i.e. it is responsible for the enantio-sensitivity of $\vec{j}$.
Using the standard expression for the amplitude of elliptically polarized field $\vec{E}_{\omega} = E_0 (\hat{x}+i\sigma\hat{y})/\sqrt{2}$, we can obtain the respective light's pseudovector: $\vec{L}=i\vert E_0 \vert^2 \sigma \hat{z}$. It is proportional to the photon spin $\sigma \hat{z}$, therefore it vanishes for linearly polarized fields ($\sigma=0$) and points in opposite directions for opposite circular polarizations ($\sigma=\pm1$). Thus,  $\vec{L}$ is responsible for the circular dichroism of $\vec{j}$. Moreover, since the observable $\vec{j}$ is perpendicular to the polarization plane, the corresponding detector must be able to distinguish between the two opposite directions perpendicular to the polarization plane, i.e. it must define a reference vector (e.g. $\hat{z}$), 
on which  the vector of photoelectron current $\vec{j}$ can be projected. Otherwise, the detector will not have the capacity of recording the circular dichroism or the enantio-sensitivity encoded in $\vec{j}$. Indeed, one can show that the coefficient $b_{1,0}$ usually measured in PECD satisfies  $b_{1,0}\propto j_z=\hat{z}\cdot\vec{j} = gS$, which is a product of the molecular handedness encoded in $g$ and handedness of the chiral setup encoded in $S\equiv\hat{z}\cdot\vec{L}$  \cite{Ordonez2018generalized}. 
 Due to the purely electric-dipole nature of PECD, the enantio-sensitive and dichroic signal  recorded in $b_{1,0}\propto j_z$ reaches several tens of percent of the total photoionization signal  \cite{Ritchie1976PRA,Powis2000JCP,Bowering2001PRL,Lischke2004,Turchini2004,Stener2004,Stranges2005,Tommaso2006,Turchini2009,Nahon2015JESRP,Ferre2015,Turchini2017,Fehre2021,Kruger2021,Garcia2013,Lux2012Angewandte,Lehmann2013JCP,Stefan2013JCP,Fanood2014,Fanood2015,Lux2015,Lux2015JPB,Beaulieu2016Faraday,Kastner2016CPC,Miles2017,Kastner2019,Ranecky2022,Comby2016JPCL,Beaulieu2016NJP,Beaulieu2018PXCD,Powis2008,Janssen2014PCCP,Hadidi2018}.
 
Importantly, the vector product of the two photoionization dipoles is equal to 
zero for the ``flat'' (plane wave) continuum:  
the formation of an enantio-sensitive electron current in the continuum
upon photoionization from a stationary state requires electron scattering from the molecular potential. This scattering imparts angular momentum on the electron, which, together with its linear momentum $\vec{k}$, characterizes chirality of the electron dynamics in the continuum, i.e. the chirality of the photoelectron current. Eq. \ref{PECD_mol_pseudoscalar} points to the physical origin of PECD and its connection to the concept of geometric fields (See Section 8.4).$\blacksquare$

\smallskip
\textbf{Rule 2:}
\textsf{A chiral observer can detect enantio-sensitive signals in the  electric-dipole approximation  due to even-order 
non-linear optical processes, such as those
described by even-order electric susceptibilities \cite{boyd_nonlinear_2008,Neufeld2019PRX,Ayuso2022OptExp}
$\chi^{(2)}$, $\chi^{(4)}$, ... }.
\smallskip

It is easy to understand this rule. Here we deal with excitation or detection of induced polarization $\vec{P}$. From \textbf{Rule 1}, we already know that any vectorial enantio-sensitive signal, such as $\vec{P}$, must be proportional to a pseudovector coming from the electromagnetic field. Let us look at all available pseudovectors. 

In the first order with respect to the field, there is only one pseudovector  $\vec{H}_{\omega}$ -- the magnetic component of EM wave.  
In the second order, we have the first opportunity to construct the field pseudovector from the vector product  of 
the electric fields,
$\left[\vec{E}_{\omega_1}\times\vec{E}^{*}_{\omega_2}\right]$. 
In the third order, to construct a field pseudovector, we have to use the magnetic field again, e.g. 
$\vec{H}_{\omega_1}(\vec{E}_{\omega_2}\cdot\vec{E}^{*}_{\omega_3})$.
Here we have used the pseudovector $\vec{H}_{\omega_1}$ and multiplied it by a scalar $\vec{E}_{\omega_2}\cdot\vec{E}^{*}_{\omega_3}$. In the fourth order, we can use the field pseudovector in the electric-dipole approximation $\left[\vec{E}_{\omega_1}\times\vec{E}^{*}_{\omega_2}\right]$ and combine it with the scalar $\vec{E}_{\omega_3}\cdot\vec{E}^{*}_{\omega_4}$. The rest is clear: in even orders, the pseudovector can be constructed from electric field vectors, but in odd orders there is no such an opportunity and we have to use the magnetic field instead. Thus, non-linear optical processes of even order can be enantio-sensitive in the electric-dipole approximation, while the enantio-sensitivity of  odd order processes relies on magnetic interactions. The same conclusion can be obtained using simple symmetry arguments   \cite{neufeld_floquet_2019, Neufeld2019PRX}.

To illustrate \textsf{{\bf Rules 1 and 2}}, we shall consider several phenomena which arise in different fields of research, but have identical mechanism of enantio-sensitive response.

\begin{figure}
\centering
\includegraphics[width=\linewidth, keepaspectratio=true]{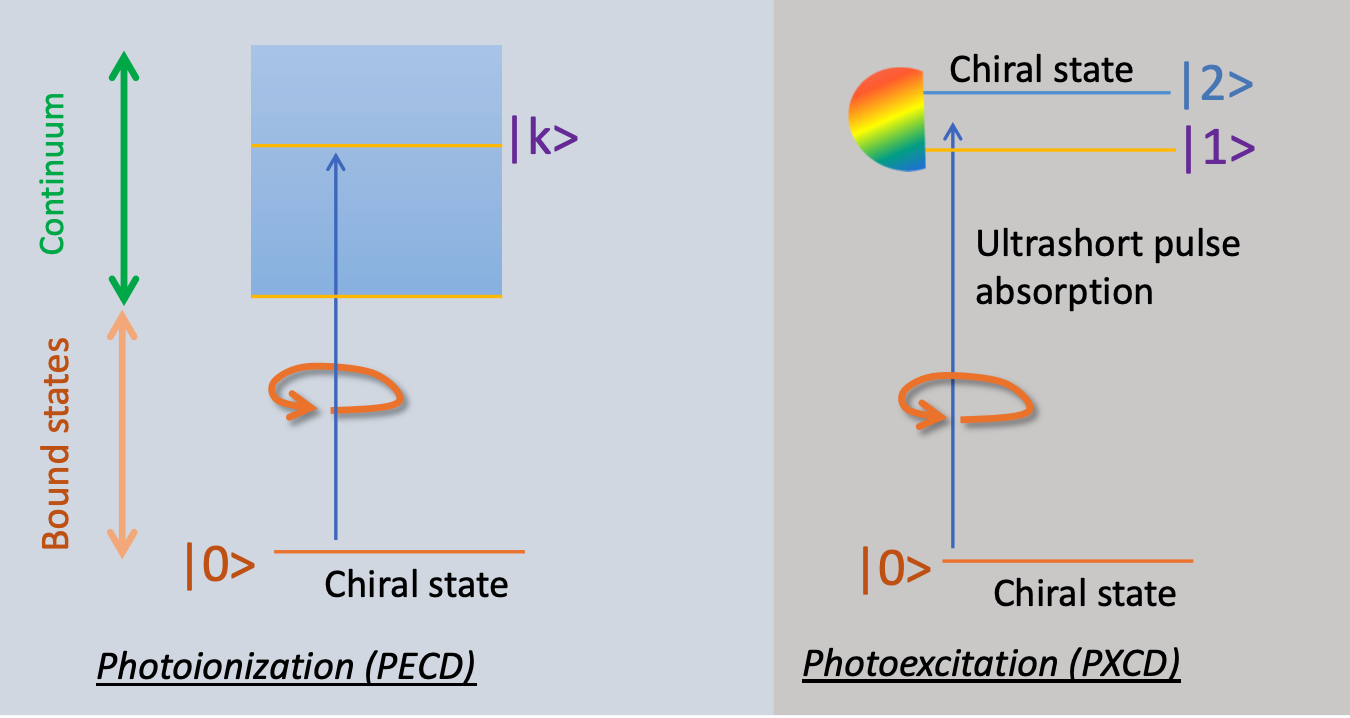}
\caption{\textbf{}
Photoelectron circular dichroism (left) vs photoexitation circular dichroism (right). In PECD circularly polarized light induces chiral  current in continuum, while in PXCD, it induces chiral current in bound states.  }
\label{fig_PXCDvsPECD}
\end{figure}

\textbf{Photo-excitation circular dichroism: helical currents in bound molecular states.}
 Consider first the phenomenon of photo-excitation circular dichroism (PXCD), introduced recently by Beaulieu et al  \cite{Beaulieu2018PXCD}. It shows that chiral photoelectron currents, which underlie the photoionization circular dichroism (PECD) discussed above, can also be excited in bound states (Fig.\ref{fig_PXCDvsPECD}). 
 
\begin{figure}
\centering
\includegraphics[width=\linewidth, keepaspectratio=true]{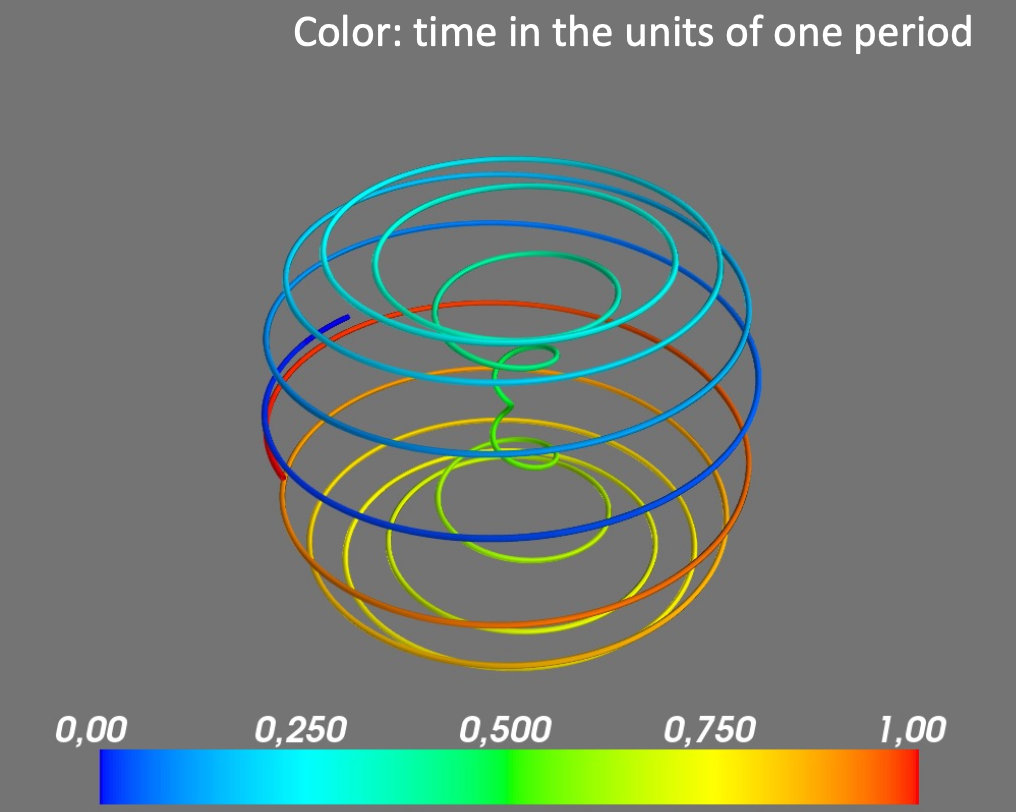}
\caption{\textbf{}
Helical current traced by the tip of the excited PXCD dipole  \cite{Beaulieu2018PXCD}.}
\label{fig_HelicalCurrent}
\end{figure}

In PXCD  \cite{Beaulieu2018PXCD, Ordonez2018generalized} a short pulse
\begin{equation}
\vec{E}(t)=\int \mathrm{d}\omega \vec{E}_{\omega}e^{-i\omega t} 
\end{equation}
interacts with an isotropic sample of chiral molecules and coherently excites a pair of vibrational or electronic states $|1\rangle$ and $|2\rangle$ via one-photon transitions. Coherent excitation of a pair of states by a circularly polarized pulse should lead to dynamics. What kind of dynamics? Recall that chiral media can convert in-plane rotation (excited by the circularly polarized field) into linear motion orthogonal to the plane.  Thus, one would expect that, after the pulse, the tip of the vector describing the induced polarization in the randomly oriented molecular ensemble traces a helical trajectory in time (Fig. \ref{fig_HelicalCurrent} ). In contrast to continuum states, in bound states this motion will get reflected from the outer turning point of the bound trajectory, reversing the direction of the motion orthogonal to the polarization plane of the exciting pump pulse. Formally, one can indeed show that, upon averaging over random molecular orientations (denoted as $\langle ...\rangle$), the dynamics in the excited states does lead to oscillations of the 
expectation value of the induced dipole $\langle \vec{d} \rangle$ in the direction perpendicular to the polarization plane  \cite{Beaulieu2018PXCD}. In the frequency domain, this dipole is:
\begin{equation}
\langle\vec{d}\rangle(\omega_{2,1})=g\vec{L},
\label{eq:gL_PXCD}
\end{equation}
where 
\begin{equation}
g\equiv \frac{1}{6}(\vec{d}_{1,0}\times\vec{d}_{2,0})\cdot\vec{d}_{2,1}
\label{eq:g_PXCD}
\end{equation}
and 
\begin{equation}
\vec{L}\equiv \vec{E}^{*}(\omega_{1,0})\times \vec{E}(\omega_{2,0}).
\label{eq:L_PXCD}
\end{equation}
That is, we observe oscillations at the difference frequency $\omega_{2,1} \equiv \omega_2 - \omega_1$ in the direction determined by $\vec{L}$. The excited dipole is indeed 
a result of the conversion of the initial in-plane rotation, excited in the electronic or vibrational degrees of freedom by the circularly polarized pump field, into motion orthogonal to this plane. The phase of these oscillations is determined not only by $\vec{L}$ but also by the sign of the molecular pseudoscalar $g$. Thus, oscillations excited in opposite enantiomers will have opposite phases. Observing an enantio-sensitive response requires detecting the phase of the 
oscillating signal, or equivalently the direction of the respective vectorial observable. 

While PXCD is similar to PECD, the conversion of 
in-plane rotation into motion orthogonal to the plane 
happens in bound states and requires at least two excited states (instead of one continuum state in PECD). Therefore, we are now dealing with the Fourier components of the field at two different frequencies. This aspect opens more options for the pump pulse polarizations which lead to a non-zero $\vec{L}$. For example, one can use an elliptically polarized broadband pulse, e.g. $\vec{E}_{\omega}=\mathcal{E}_{\omega}(\hat{x}+i\sigma\hat{y})/\sqrt{2}$, which yields $\vec{L}=i\mathcal{E}^*_{\omega_{1,0}}\mathcal{E}_{\omega_{2,0}}\sigma \hat{z}$, or two linearly polarized pulses at an angle with respect to each other, e.g. $\vec{E}_{\omega_{1,0}} = \mathcal{E}_{\omega_{1,0}}\hat{x}$ and  $\vec{E}_{\omega_{2,0}} = \mathcal{E}_{\omega_{2,0}}\hat{y}$, which yields $\vec{L}=\mathcal{E}^*_{\omega_{1,0}}\mathcal{E}_{\omega_{2,0}}\hat{z}$. $\blacksquare$

The PXCD signal is an attractive new enantio-senstive molecular observable. It depends on molecular properties (dipoles) and hence it is molecule specific. It relies on coherence between the states and, hence, it tracks coherence, electronic or vibronic. 
For example, in the experiment  \cite{Beaulieu2018PXCD} such vibronic coherence,  excited in Rydberg states of fenchone and camphor by a circularly polarized femtosecond pulse, gave rise to chiral vibronic currents lasting for 1.2 ps. Its evolution reflected significant variations of molecular chirality when probed via photoionization (see PXECD below). Finally, since the emission associated with 
the PXCD dynamics is a parametric process, signals from all molecules in the ensemble will add coherently to form macroscopic dipole. This opens an opportunity to control enantio-sensitive response not only on a single molecule level, but also at the macroscopic level. We will consider such 
macroscopic  control of chiral emitters later, 
developing the concept of light with structured chirality  in Section 5.  

\textbf{Enantio-sensitive microwave spectroscopy.}
A phenomenon analogous to PXCD exists in the microwave domain. Known as enantio-sensitive microwave spectroscopy (EMWS), it occurs in the case of purely rotational transitions,  was 
discovered by Patterson et al \cite{Patterson2013PRL} and recently further developed in  \cite{Yachmenev2016PRL,owens2018climbing,Gershnabel2018PRL,tutunnikov2020observation,Yachmenev2019PRL,Milner2019PRL}.  Indeed, as shown in Ref.  \cite{Ordonez2018generalized}, the rotational problem can be formulated in a way mathematically similar to PXCD in 
electronic or vibrational states, with the only difference that 
the expression for the molecular pseudoscalar includes a sum over all magnetic quantum numbers $M_i$ of the asymmetric rotor states in each energy level $i=0,1,2$: 
\begin{equation}
g\equiv\frac{1}{6}\sum_{M_0,M_1,M_2} (\vec{d}_{1M_1,0M_0}\times\vec{d}_{2M_2,0M_0})\cdot\vec{d}_{2M_2,1M_1}, 
\end{equation}
This difference stems only from the different nature of averaging over molecular orientations in PXCD and EMWS  \cite{Ordonez2018generalized}.   In   averaging over molecular orientations, the  coherence of different rotational states is not important in PXCD, because it does not require coherent excitation of rotational states. One can view different molecular orientations as  different molecules (of the same kind) and perform classical averaging over molecular orientations  \cite{Ordonez2018generalized}. In the case of EMWS, coherent excitation of rotational states is directly involved and the analysis requires quantum averaging   \cite{Ordonez2018generalized}.$\blacksquare$

\textbf{Enantio-sensitive non-linear wave-mixing.}
PXCD and EMWS are not the only two relatives in the family of enantio-sensitive phenomena which occur in the electric-dipole approximation and follow \textbf{Rule 1}.
Although PXCD and EMWS are resonant processes, and 
oscillations take place in the absence of the driving field, the pseudoscalar $g$ can loosely be interpreted as a second order susceptibility $\chi^{(2)}$. Indeed, PXCD and EMWS are closely related to the original predictions of Giordmaine  \cite{Giordmaine1965} for sum- and difference-frequency generation in chiral liquids. The expression for the second-order induced polarization at, e.g. the difference frequency can be rewritten in the form dictated by our \textbf{Rule 1}: 
\begin{equation}
\vec{P}^{(2)}(\omega_{2,1})=\chi^{(2)}\vec{L},
\end{equation}
The non-linear susceptibility $\chi^{(2)}$ is a sum over states $\vert i\rangle$ and $\vert j\rangle$ involving terms of the form $(\vec{d}_{i,0}\times\vec{d}_{j,0})\cdot\vec{d}_{j,i}$; $\vec{L}$ is given by Eq. (\ref{eq:L_PXCD}).

\begin{figure}
\centering
\includegraphics[width=\linewidth, keepaspectratio=true]{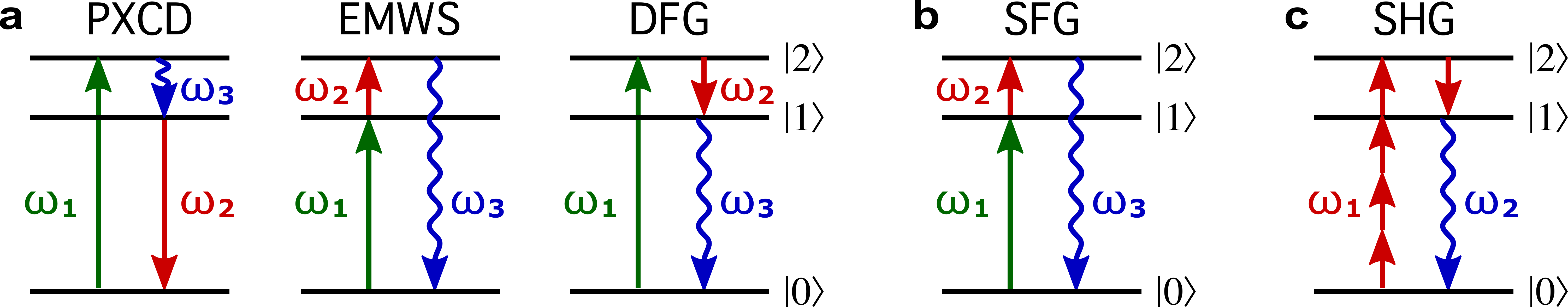}
\caption{\textbf{Diagrams for various nonlinear  enantio-sensitive or chiral-sensitive processes. Different colors mark different directions of light polarization (e.g. "green" -$x$, "red" - $y$, "blue" - $z$).} 
\textbf{a} Diagrams for photo-excitation circular dichroism (PXCD), difference-frequency generation, and enantio-sensitive microwave spectroscopy (EMWS);
\textbf{b} Sum-frequency generation (SFG). If $\omega_1=\omega_2$, SFG can not be enantio-sensitive, because the respective spatial directions become non-distinguishable. It means that second harmonic generation (SHG) is prohibited in randomly oriented chiral molecules in the lowest-order perturbative regime  \cite{fischer_chiral_2002}.
\textbf{c} Second harmonic generation becomes enantio-sensitive in elliptically polarized field $\omega$ in the non-perturbative regime, because the respective diagram records the direction of field rotation via one up and one down arrow.}
\label{fig_NLO_Diagrams}
\end{figure}

Fig \ref{fig_NLO_Diagrams}  shows non-linear optics-type diagram for  PXCD, EMWS, (Fig \ref{fig_NLO_Diagrams}a) and difference frequency generation (DFG) (Fig \ref{fig_NLO_Diagrams}b), 
which unifies all three processes. The arrows directed up 
describe amplitudes of photon absorption, while the 
arrows directed down describes conjugated amplitudes 
corresponding to photon emission. These 
amplitudes are excited by the respective (conjugated or not) components of the incident field: the conjugated field is associated with the arrow directed down. 

Further advances in enantio-sensitive perturbative non-linear optics are described in the review by Fisher et al  \cite{Fischer2005}.

Similarly, in the context of highly non-linear interactions, symmetry arguments \cite{neufeld_floquet_2019,Neufeld2019PRX,Ayuso2022OptExp} show that the interaction between a periodic (but not necessarily monochromatic) electric field and an isotropic chiral medium can lead to enantio-sensitive polarization at even multiples of the fundamental frequency $\omega$ of the field. That is, while $\vec{P}^{(2n+1)\omega}$ is identical for both enantiomers, $\vec{P}^{2n\omega}$ has opposite signs for the opposite enantiomers. Analogously to the perturbative case, this means that there are electric field configurations for which even-order high harmonics are possible only if the medium is chiral \cite{Neufeld2019PRX,Ayuso2022OptExp}. 
Since in the non-perturbative regime  second harmonic generation can involve additional up-down arrows corresponding to virtual emission/absorption of photons, Fig. \ref{fig_NLO_Diagrams} c, it can also become enantio-sensitive.$\blacksquare$

In all of these methods, the ability to distinguish between opposite enantiomers requires access to the phase of the induced polarization. Thus, intensity measurements of the even order polarizations can distinguish chiral media from achiral media or different chiral media from each other, but cannot distinguish between opposite enantiomers.

In the next sub-section (see cHHGd), we shall see how enantio-sensitive information can be mapped on the polarization ellipse of the nonlinear optical response. 

\textbf{Chiral high harmonic generation in the electric-dipole approximation (cHHGd).}
High harmonic generation (HHG) is an extremely nonlinear process that converts intense radiation, usually in the IR domain, into high-energy photons with frequencies that are high-integer multiples of the incident field frequency \cite{Krausz2009RMP}. It can be understood semi-classically as a sequence of three steps, starting with strong-field ionization  \cite{Corkum1993}. In the second step, the laser electric field takes the liberated electron away from the core, driving its
oscillations in the continuum. Note that the laser field can also interact with the ionic core, driving rich multi-electron dynamics  \cite{Smirnova2014Wiley}. The third step is the electron-core recombination, resulting in the emission of the harmonic light. A typical HHG spectrum contains information about both the structure of the atomic/molecular system and the ultrafast dynamics between ionization and recombination.

\begin{figure}
\centering
\includegraphics[width=\linewidth, keepaspectratio=true]{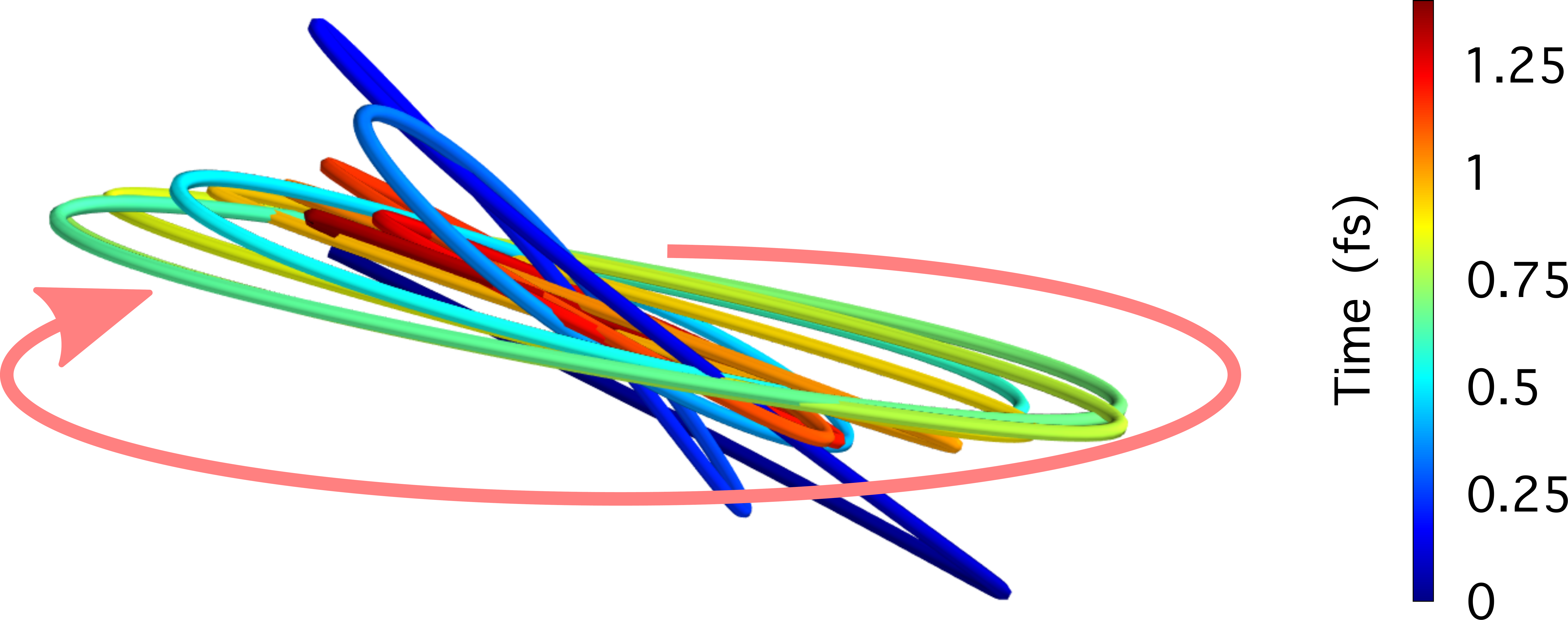}
\caption{Time-dependent polarization driven in randomly oriented propylene oxide by an elliptically polarized field with intensity $5\cdot10^{13}$W/cm$^2$, wavelength 1770nm and 5$\%$ of ellipticity (see  \cite{Ayuso2022OptExp} for details of the calculations).
The polarization of the driving field is depicted in pink.
The ultrafast polarization response is 3D, chiral, and enantio-sensitive: the in-plane (achiral) polarization components are identical in left- and right-handed molecules, whereas the (chiral) out-of plane component is out of phase in opposite enantiomers (compare with the helix in Fig. \ref{fig_HelicalCurrent}). }
\label{fig_3Ddipole}
\end{figure}

The third step of HHG, radiative electron-core recombination, is the inverse of photoionization. Therefore, just like in the photoionization in PECD, the recombination dipole responsible for HHG driven by an elliptically polarized field in isotropic chiral media can develop a component which is orthogonal to the polarization of the driving field, i.e. along the direction of light propagation, as show in Fig. \ref{fig_3Ddipole}.
This component oscillates out of phase in media of opposite handedness. However, unlike the photoelectron current in PECD, this chiral dipole component escapes direct observation in standard HHG measurements: the dipole oscillating along the light propagation direction radiates orthogonal to it. This means that in the electric-dipole approximation the chiral component propagates in the direction orthogonal to the direction of the driving field and thus cannot be observed in the macroscopic HHG signal. Indeed, the macroscopic HHG signal requires coherent addition of the light emitted from different positions in the chiral medium, which is only possible if the harmonic light co-propagates with the driver. Indeed, in the first chiral HHG experiments \cite{Cireasa2015} and subsequent setups  \cite{Smirnova_2015JPB,Ayuso2018JPB, Ayuso2018JPB_model,Baykusheva2018PRX} this enantio-sensitive dipole component could not be observed, and the enantio-sensitive response relied on the interaction of chiral molecules with the magnetic component of the laser field.

The fact that the electric-field vector of the intense laser field driving HHG is orthogonal to its propagation direction stops us from imprinting the strong chiral response shown in Fig. \ref{fig_3Ddipole} into the macroscopic HHG signal, severely limiting the potential of the HHG camera.
This problem can be overcome by creating light with ``forward'' ellipticity, or transverse spin \cite{Bliokh2015}, and using to drive chiral HHG.

\begin{figure}[h]
\centering
\includegraphics[width=\linewidth, keepaspectratio=true]{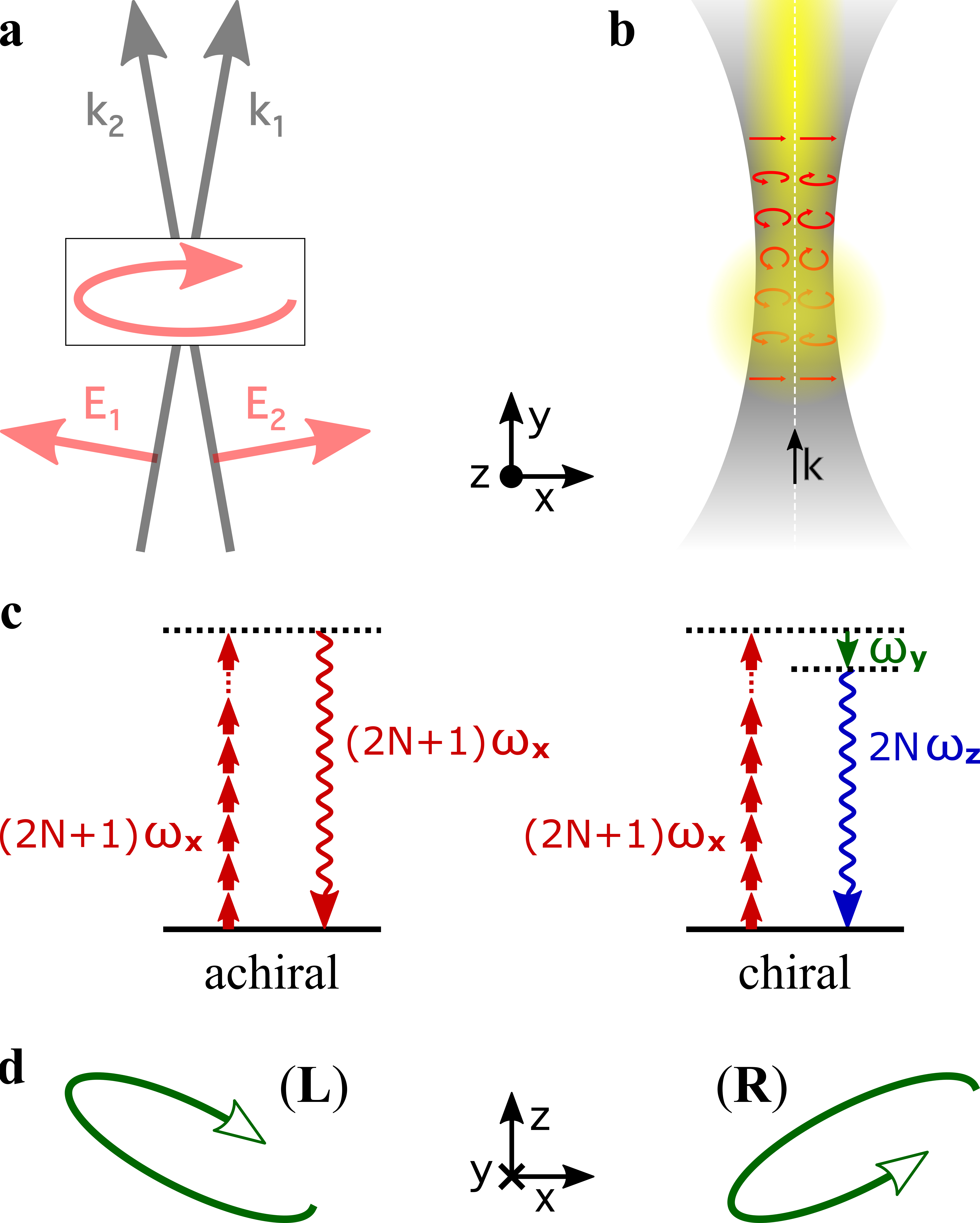}
\caption{\textbf{Driving chiral HHG driven using light with ``forward'' ellipticity.}
\textbf{a} Schematic representation of a non-collinear setup for creating a field with ``forward'' ellipticity;
\textbf{b} A Gaussian beam acquires a strong longitudinal component upon tight focusing, also leading to ``forward ellipticity'', which has opposite sign on the opposite sides of the laser beam axis.
\textbf{c} Achiral response corresponds to odd harmonics, chiral  response corresponds to even harmonics and is out of phase in two enantiomers. \textbf{d} Spectral overlap of odd-order 
and even-order nonlinear-optical response in a few cycle pulses  leads to opposite rotation of the polarization ellipse in opposite enantiomers.
}
\label{fig_One-color-Setup}
\end{figure}

Consider two laser beams that propagate non-collinearly at a small angle, both carrying the same fundamental frequency and  linearly polarized in the propagation plane. In the overlap region, the total electric field becomes elliptically polarized in the plane of propagation ($x,y$), with the minor ellipticity component in the propagation direction ($y$-axis), see Fig. \ref{fig_One-color-Setup}a.
Now the chiral dipole component has adequate orientation to generate harmonic light that co-propagates with the driving field. Thus, it can be mapped onto the macroscopic HHG signal  \cite{Ayuso2022OptExp}.

Non-collinear optical setups are not the only way of creating electric-field vector components along the propagation direction which lead to forward ellipticity.
Such longitudinal components naturally arise when light is confined in space, as it happens in a tightly focused laser beam, see Fig. \ref{fig_One-color-Setup}b. 

The HHG signal driven by light with forward ellipticity has two orthogonally polarized components: the (standard) component in the propagation plane, which is not sensitive to chirality, and the enantio-sensitive component orthogonal to it, Fig. \ref{fig_One-color-Setup}b, which is out of phase in media of opposite handedness. One might think that the harmonic light would be elliptically polarized, but this is not 
necessarily the case. Indeed, the \textsf{{\bf Rule 2}} dictates that the enantio-sensitive and non-enantio-sensitive components of the induced polarization carry different harmonic frequencies: the non-enantio-sensitive component carries odd harmonics, the enantio-sensitive component carries even harmonics, Fig. \ref{fig_One-color-Setup} c, (see \textbf{Rule 1}). Hence, for relatively long pulses, the non-collinear setup in Fig \ref{fig_One-color-Setup} generates non-overlapping even and odd harmonics, which are orthogonally polarized. Only the phase of the even harmonics records the medium's handedness \cite{Ayuso2022OptExp} as it follows from \textsf{{\bf Rules 1, 2}}. But, what if the driving pulse is only a few cycles long, and therefore has broad spectrum?

\textbf{Nonlinear optical rotation.}
In chiral HHG driven by a few-cycle laser pulse which has forward ellipticity, the chiral (even-order) and achiral (odd-order) harmonics can spectrally overlap. Then the emitted harmonic light becomes elliptically polarized \cite{Ayuso2021Optica} in the overlap region.  Since the chiral components of the nonlinear-optical response are out of phase in opposite enantiomers, both the spin (ellipticity) and
the rotation of the polarization ellipse of the 
nonlinear-optical response in the spectral overlap region 
will be opposite in opposite enantiomers. 

Fundamentally, using the ultra-broad spectrum of a tightly focused few-cycle pulse \cite{Ayuso2021Optica} is akin to using a two-colour non-collinear setup.
Such a setup allows one to tailor the polarization of the driver in two and three dimensions and maps the molecular handedness onto the polarization properties of the emitted harmonic light.

A polarizer placed before the high harmonic detector and rotated by an angle $0<\alpha<90^\circ$ with respect to the linearly polarized driving field can convert the enantio-sensitive orientations of the harmonic polarization ellipse into different signal intensities \cite{Ayuso2021Optica}. 
Just like the traditional optical activity, its nonlinear analogue does not require chiral light (see Section 2.1.1).
As a result, the total number of photons that reach the detector does not depend on the medium's handedness.
Note that the rotation of the polarization ellipse of the emitted light \cite{Ayuso2021Optica} is analogous to the rotation of the linear polarization in standard optical rotation. Indeed, in the non-linear optical rotation \cite{Ayuso2021Optica}, the chiral setup is formed by the pseudovector defining rotations of the polarization ellipse and by the vector defining the propagation direction $\hat{k}$ of the light. $\hat{k}$ is relevant in this electric-dipole effect not only because of phase-matching conditions along $\hat{k}$ but also because $\hat{k}$ is encoded in the pattern of forward ellipticities of the driving field, a phenomenon known as transverse spin-momentum locking  \cite{Bliokh2015}.

\smallskip
\textbf{Rule 3.}
\textsf{A chiral observer allows the detection of enantio-sensitive tensorial observables, e.g. quadrupolar currents and quadrupolar polarizations.%\ \texttt{\textbackslash 
\footnote{Note that here and below the term \emph{quadrupolar} refers to the structure of the observable itself and is unrelated to effects emerging from electric-quadrupole interactions. The effects discussed here are always within the electric-dipole approximation.}}
\smallskip

Tensorial observables are particularly relevant in two-color fields, such as a laser beam carrying $\omega$ and $2\omega$ frequencies, arranged so that enantio-sensitivity at the level of vectorial observables is symmetry forbidden. The two-color aspect of these fields naturally brings perspectives of coherent control into the discussion. Let us consider some examples.

\textbf{Photoionization in two-color fields without spin.}
The light pseudovector in many electric-dipole-based methods is proportional to the fields's spin. In the absence of spin, there can be no vectorial enantio-sensitive observables. For example, laser beams carrying $\omega$ and $2\omega$ frequencies, linearly polarized and orthogonal to each other, do not carry spin: their vector product changes direction during the laser cycle and is equal to zero on average. Hence, they cannot excite vectorial enantio-sensitive observables in randomly oriented ensembles. Yet,  enantio-sensitive photoionization signals triggered by such fields have been predicted  \cite{Demekhin2018PRL,Demekhin2019PRA} and detected  \cite{Rozen2019PRX}.

The point here is that, in addition to vectorial observables such as induced polarization (ESMW, PXCD, difference- and sum-frequency generation, cHHGd), photoelectron current (PECD, PEXCD), chiral measurements can also yield tensorial observables. The two-color arrangement above takes advantage of such observables. 

A simple way to introduce tensorial observables is to 
consider photoionization \cite{ordonez_molecular_2020}. Indeed, any photo-electron angular distribution $W(\theta,\phi)$ can be decomposed in spherical harmonics as
\begin{equation}
W(\theta,\phi)=\sum_{l,m} b_{l,m} Y_{l,m}(\theta,\phi), 
\end{equation}
where the values of the coefficients $b_{l,m}$ for fixed $l$ and $-l\leq m \leq l$ are the entries of a spherical tensor of rank $l$ and dimension $(2l+1)$  \cite{brink_angular_1968}. The determination of these coefficients is therefore equivalent to a tensorial measurement. 

Since vectors are tensors of rank one, for $l=1$ we obtain the vectorial observables we have already discussed. In PECD, for example, where $W(k,\theta,\phi)$ is the photoelectron angular distribution at photoelectron energy $k^2/2$,  
the three coefficients $b_{1,m}$ describe the 
vector of the photoelectron current; $b_{1,\pm1}=0$ due to symmetry and $b_{1,0}\neq0$ is responsible for the so-called forward-backward asymmetry. In the bound case, e.g. in PXCD, one could employ $W(r,\theta,\phi)$ to describe bound electron density. In this case the $b_{1,m}$ coefficients describe the expectation value of the electric-dipole operator coupling the two states excited by the pump pulse.

Just like the  $b_{1,m}$ coefficients encode vectorial properties,  
the $b_{2,m}$ coefficients describe the quadrupolar part of $W(\theta,\phi)$. 
In the context of photoionization with circularly polarized light, it is well known that these coefficients are not enantio-sensitive. However, they become enantio-sensitive and correspond to excitation of an enantio-sensitive quadrupolar photoelectron current resulting from the interference between a two-$\omega$-photon and a one-$2\omega$-photon ionization pathway when the $\omega$ and $2\omega$ fields are linearly polarized perpendicular to each other, as predicted in  \cite{Demekhin2018PRL,Demekhin2019PRA}. These predictions have been supported by the observation of similar enantio-sensitive multipolar signals with $l\geq2$ in strong-field ionization of fenchone and camphor \cite{Rozen2019PRX}. The analysis of the symmetry properties of this electric field configuration reveals that, together with an appropriately oriented quadrupolar detector, it yields a chiral setup (see Fig. \ref{fig_scheme}c), and its handedness emerges naturally in the expressions for the  enantio-sensitive response \cite{ordonez_disentangling_2020}. 

Let us consider interference of  the one-photon ionization pathway of the initial state $|0\rangle$ triggered by the $2\omega$-field with the two-photon ionization pathway via
an intermediate state $|1\rangle$, triggered by the $\omega$ field. The fields are defined as
\begin{equation}
\vec{E}(t)=\vec{E}_{\omega}e^{-i\omega t} + \vec{E}_{2\omega}e^{-2i\omega t} + \mathrm{c.c.},
\end{equation}
with $\vec{E}_{\omega}=E_{\omega}\hat{x}$  and $\vec{E}_{2\omega}=E_{2\omega}e^{-i\phi}\hat{z}$. One can show that the coefficient describing an $xy$ quadrupole photoionization yields is \cite{ordonez_disentangling_2020}
\begin{equation}
\tilde{b}_{2,-2}=A^{(1)*}A^{(2)}gS+\mathrm{c.c.},
\end{equation}
where tilde indicates that the expansion was over the real spherical harmonics, the coefficients $A^{(1)}$ and $A^{(2)}$  depend on the detunings and pulse envelopes,
\begin{equation}
g=\int \mathrm{d}\Omega_k \{[\hat{k}\cdot(\vec{d}_{\vec{k},0}\times\vec{d}_{\vec{k},1})](\hat{k}\cdot \vec{d}_{1,0}) + [\hat{k}\cdot(\vec{d}_{\vec{k},0}\times\vec{d}_{1,0})](\hat{k}\cdot\vec{d}_{\vec{k},1}) \}
\end{equation}
is a complex-valued molecular pseudoscalar with a structure analogous to that found in PECD, and
\begin{equation}
S=(\vec{E}_{\omega}\cdot\vec{E}_{\omega})[\vec{E}_{2\omega}^{*}\cdot(\hat{x}\times\hat{y})]
\end{equation}
is a pseudoscalar that encodes the handedness of the chiral setup. Note the emergence of two axes, $\hat{x}$ and $\hat{y}$, in the expression for $S$. They highlight the role of the detector (see Fig. \ref{fig_scheme}c) in defining a reference frame that distinguishes directions for which the product $xy$ is positive from directions for which it is negative (see also discussion related to Fig. \ref{fig_scheme}). Note that $S$ depends on the two-color phase $\phi$, which can be used as a control parameter. The phase $\phi$ is analogous to the relative phase between $\vec{E}^*(\omega_{1,0})$ and $\vec{E}(\omega_{2,0})$ in PXCD and to the relative phase between the two perpendicular components of a circularly polarized light in PECD.

Just like vectorial enantio-sensitive electron current in photoionization (PECD) has its counterpart in bound states (PXCD), the quadrupolar current found in photoionization  \cite{Demekhin2018PRL, Demekhin2019PRA,Rozen2019PRX} has a quadrupolar analogue in the context of bound excitation. That is, the same field configuration and the same interference scheme translated to the context of bound excitation leads to the emergence of an enantio-sensitive quadrupole \cite{ordonez_inducing_2021}. There is, however, an important difference from  the induced bound-state dipole in PXCD: the intereference here arises from the two pathways, one-photon and two-photon, and no longer requires coherent population of two electronic or vibronic states: a single final bound electronic state is sufficient. Consequently,  the generated quadrupole is permanent: it does not oscillate 
and is associated with uniaxial orientation of the (initially isotropic) molecular sample. $\blacksquare$

Note that, in all phenomena considered in this section, the total intensity of the nonlinear-optical or photoelectron signal is not enantio-sensitive because the driving fields are not chiral.
Achieving enantio-sensitivity in the total signal intensity
 requires an efficient chiral photonic reagent: synthetic chiral light. 

\section{The second electric-dipole revolution in chiral measurements: Efficient Chiral Reagent}\label{sec:chiral_reagent}

\begin{figure}[h]
\centering
\includegraphics[width=\linewidth, keepaspectratio=true]{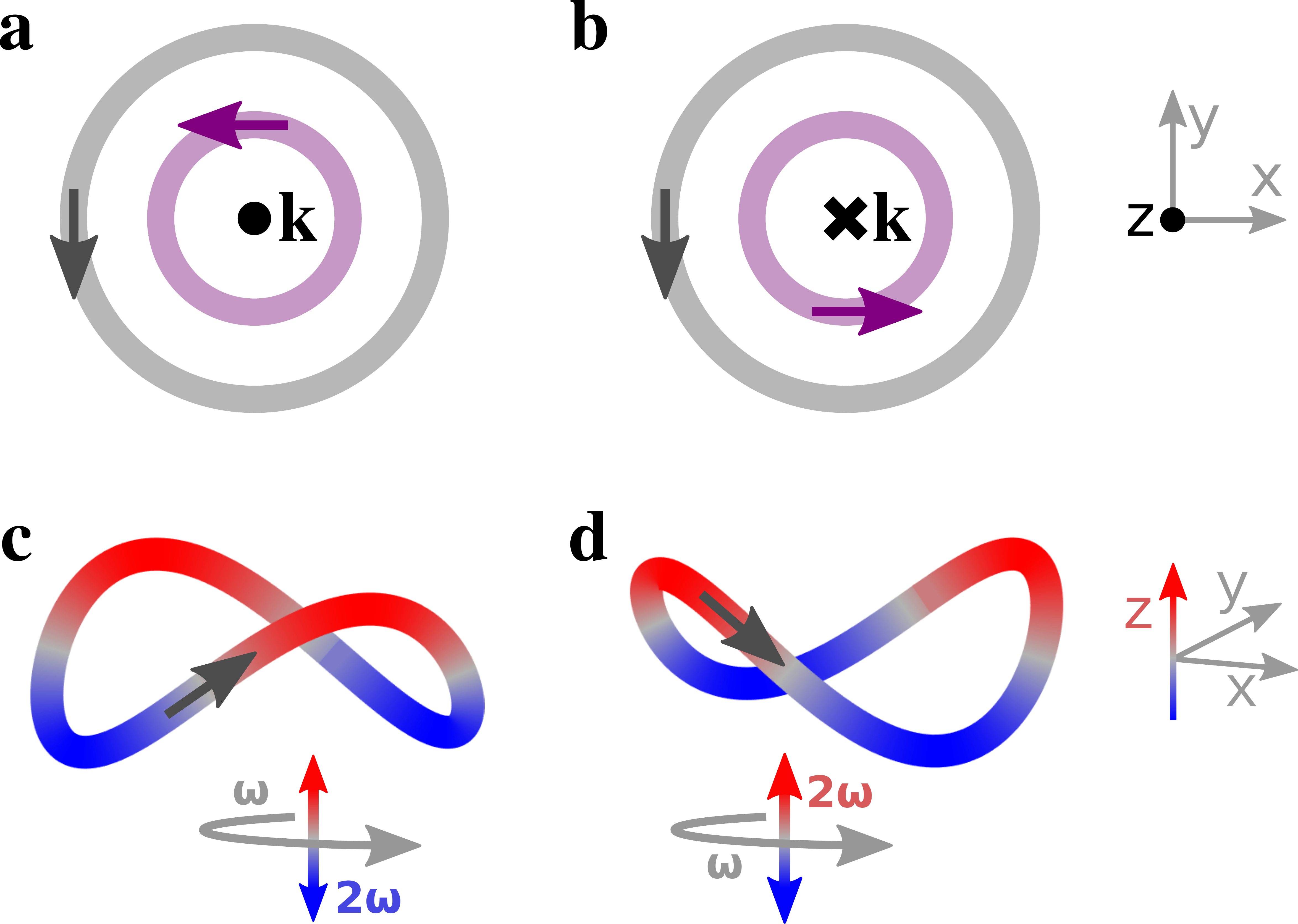}
\caption{\textbf{The concept of synthetic chiral light.}
\textbf{a}, Electric field vector (red arrow) of circularly polarized light draws helix as light propagates in space. \textbf{b} Synthetic chiral light is locally chiral, because the tip of its electric field vector (red arrow) draws chiral Lissajous figure at every point in space. Colour indicates temporal evolution of the trajectory, drawn by the tip of the electric field vector.
}
\label{fig_CPLvsSCL}
\end{figure}

A chiral reagent provides access to fundamentally different enantio-sensitive observables upon interaction with chiral matter: total enantio-sensitive intensity signals, not accessible by a chiral observer. Circularly polarized light, the standard chiral photonic reagent, owes its handedness to the (chiral) helix that the electric-field vector draws in \emph{space} (see Fig. \ref{fig_CPLvsSCL}a), which can either be left- or right-handed. Its chirality is, however, non-local --at any given point in space, the electric-field vector draws a planar circle.
A fundamentally different way of endowing light with chirality is to encode it in \emph{time}, making the trajectory that the tip of the electric field vector traces in time chiral  \cite{Ayuso2019NatPhot}. 
In contrast to the (standard) handedness of circularly polarized light, this new type of chirality is defined \emph{locally}, at each point in space.
It arises already in the electric-dipole approximation.

\textbf{Concept 1.}
\textsf{\textbf{Locally chiral light} is chiral within the electric-dipole approximation: the tip of the electric field vector draws a (three-dimensional) chiral Lissajous figure in time.
The generation of this light requires three orthogonal polarization components and, at least, two colours.}
\smallskip

An example of a locally chiral field \cite{Ayuso2019NatPhot} is shown in Fig. \ref{fig_CPLvsSCL}b.
The combination of a fundamental field, which is elliptically polarized in the propagation ($xy$) plane, and an orthogonally polarized second harmonic generates chiral Lissajous figure.
Indeed, if we reflect field's trajectory, e.g.  through the $xy$ plane, the elliptically polarized $\omega$-field remains the same, but the $z$-polarized $2\omega$ component flips sign.
These mirror-reflected Lissajous figures cannot be superimposed by any rotation and/or translation.

Such locally chiral light can drive strongly enantio-sensitive optical signals in isotropic chiral matter via purely electric-dipole interactions.
Control over the temporal structure of the light field should enable efficient control over the enantio-sensitive response of chiral matter \cite{Ayuso2019NatPhot}.
For the field presented in Figs. \ref{fig_CPLvsSCL}b, three key parameters enable such control (for a given total intensity): the ellipticity of the fundamental field, the amplitude of the second harmonic, and the two-colour phase delay.

As we know from Section 2, a chiral ``object'', such as  locally chiral light, must have at least one pseudoscalar which characterizes its handedness. What is the chirality measure (pseudoscalar) of this locally chiral light?  

\noindent\textbf{Concept 2. }
\textsf{\textbf{Chiral correlation functions}
characterize  the local handedness of synthetic chiral light by recording the correlated interplay between the different frequency components of the light wave, which encodes the handedness of the Lissajus figure. Chiral correlation functions characterize the strength of non-linear enantio-sensitive light-matter interaction in the electric-dipole approximation.}

To characterize the handedness of light's Lissajous figure, one can take three  snapshots of the electric-field vector $\vec{F}(t)$ at three successive instants of time ${t_1}$, ${t_2}$ and ${t_3}$, 
and construct a triple product of these three vectors $ \vec{F}(t_1) \cdot [\vec{F}(t_2)\times\vec{F}(t_3)]$. If such a product is non-zero, it means that between $t_1$ and $t_3$ the tip of the electric field vector traced a chiral trajectory in the space of $F_x,F_y,F_z$.
To make sure that not only a section, but the entire  Lissajous curve is chiral, one can average the triple product over time:
\begin{equation}\label{eq_H3}
H^{(3)}(\tau_1,\tau_2) = \int_0^T dt \vec{F}(t) \cdot [\vec{F}(t+\tau_1)\times\vec{F}(t+\tau_2)] \ \ \ ,
\end{equation}
with $H^{(3)}(\tau_1,\tau_2)$ being the third-, and the 
lowest-order chiral field correlation function \footnote{Interestingly, the overall chirality of the helical trajectory  traced by the induced dipole in PXCD 
(see Fig. \ref{fig_HelicalCurrent}) is evident, since the two helices of opposite chirality inner and outer have different size. It means that, in the near-field, the light generated 
by this dipole via the free induction decay at three frequencies would be chiral in the electric-dipole approximation.}. 
The use of chiral correlation functions in the frequency domain, evaluated at the field's frequencies, is often more convenient than the direct application of time-domain expressions.
First, it removes the arbitrary choice of $\tau_1$, $\tau_2$, etc. 
Second, it provides a clear connection with the multi-photon processes that record the field's handedness and its interaction with chiral matter.
In the frequency domain, the  chiral correlation function $h^{(3)}$ is simply a triple product involving three frequency components of the laser field, $\omega_1$, $\omega_2$, and $-(\omega_1+\omega_2)$ \cite{Ayuso2019NatPhot}:
\begin{equation}\label{eq_h3w}
h^{(3)}(-\omega_3,\omega_1,\omega_2) = \vec{F}^{*}_{\omega_3} \cdot [\vec{F}_{\omega_1}\times\vec{F}_{\omega_2}]. 
\end{equation}
This triple product is non-zero if these three frequency components are non-coplanar.

\smallskip
\textbf{Rule 4.}
\textsf{The scalar enantio-sensitive response of chiral matter to locally  chiral light is nonlinear and results from the interference between a chiral even-order process and an achiral odd-order process. Thus, it involves an odd number of photons, leading to chiral light correlation functions of odd order.} \smallskip

\begin{figure}[h]
\centering
\includegraphics[width=\linewidth, keepaspectratio=true]{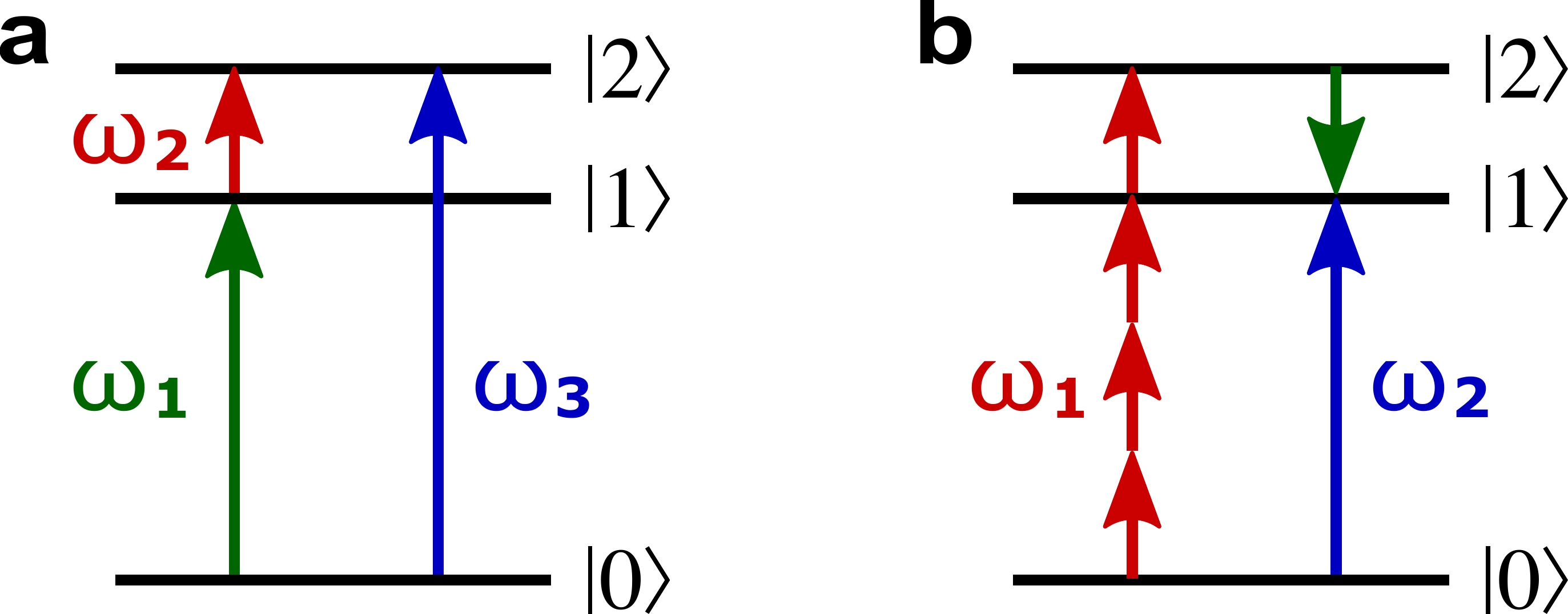}
\caption{\textbf{Enantio-sensitivity in absorption.}
Absorption occurs in a three-level system driven by fields with
frequencies $\omega_1$, $\omega_2$, and $\omega_3$ polarized along $x$, $y$, and $z$, respectively. 
The lack of inversion symmetry in a chiral molecule allows for
dipole couplings between all states.
The second-order (two-photon induced) polarization at 
$\omega_3=\omega_1+\omega_2$ is generated along $z$ in
randomly oriented chiral media.
}
\label{fig_ThreeLevelH3}
\end{figure}

For example,  $h^{(3)}$ characterizes the interference of two pathways leading to absorption in a 3-level system with chiral states 0, 1 and 2. The first (achiral) pathway is associated with the linear response at frequency $\omega_3$, which is not sensitive to chirality, and leads to polarization at $\omega_3=\omega_1+\omega_2$ along $z$.
The second (chiral) pathway corresponds to sum-frequency generation, a second-order process that records the molecular handedness: the medium absorbs one $\omega_1$ photon and one $\omega_2$ photon from
the field components polarized in the $x-y$ plane, generating polarization at frequency $\omega_3$ along $z$. 
Sum-frequency generation is symmetry-forbidden
in a randomly oriented ensemble of achiral molecules  \cite{Giordmaine1965}.
That is, this second pathway 
is unique to chiral media, and the induced polarization is out of phase in media of opposite handedness.
The two pathways interfere, making absorption and emission at frequency $\omega_3$ strongly enantio-sensitive (Fig. \ref{fig_ThreeLevelH3}). In a randomly oriented ensemble of chiral molecules, the enantio-sensitive contribution to absorption is $\Im[\chi^{(2)}h^{(3)}]$. The physical meaning of $h^{(3)}$ is clear from Table \ref{tbl:example2}, comparing standard absorption CD and non-linear absorption CD in the electric-dipole approximation: $h^{(3)}$ plays the role of optical chirality, which characterizes the strength of enantio-sensitive absorption in the linear response  \cite{Tang2010}. Likewise, in the non-linear regime, the molecular pseudoscalar  formed by the triple product of the three relevant dipoles  replaces the one typical for the linear response -- the scalar product between the electric and magnetic dipoles. 
\begin{table*}
\small
  \caption{\ Absorption CD for two types of chiral light.}
  \label{tbl:example2}
  \begin{tabular*}{\textwidth}{@{\extracolsep{\fill}}lllll}
    \hline
    Type of light & Absorption CD & Molecular psudoscalar & Light pseudoscalar\\
    \hline
    Natural light \\ (helix in space)& $\Im\{\chi_{em}OC\}$ \cite{Tang2010} & in resonance $\chi_{em}\propto[\vec{d}_{f,i}\cdot\vec{m}_{f,i}]$ & $OC\propto[\vec{E}_{\omega}^{*}\cdot\vec{B}_{\omega}]$ \\    
    \hline
    Synthetic light \\ (helix in time) & $\Im\{\chi^{(2)}h^{(3)}\}$& in resonance $\chi^{(2)}\propto[\vec{d}_{2,0}\cdot(\vec{d}_{2,1}\times\vec{d}_{1,0})]$  \cite{Fischer2000PRL} & $h^{(3)}\propto\{\vec{E}^{*}(\omega_{2,0})\cdot[\vec{E}(\omega_{2,1})\times\vec{E}(\omega_{1,0})]\}$  \cite{Ayuso2019NatPhot}  \\
    \hline    
      \end{tabular*}
\end{table*}

\textbf{Synthetic chiral light with two colors} For a two-colour field, such as the one in Fig. \ref{fig_CPLvsSCL}c, $h^{(3)}=0$ simply because the field does not contain three frequencies. 
It means that nonlinear 3-photon processes driven by this field are not enantio-sensitive, but it does not necessarily mean that the field is achiral.
If the field is locally chiral, its handedness can be recorded in higher-order processes, which are characterized and controlled by higher-order correlation functions.
They  involve additional dot products of the electric field vectors, evaluated at different times. 
The next-order chiral correlation function is
\begin{align}
\label{eq_H5}
H^{(5)}(\tau_1,\tau_2,\tau_3,\tau_4) = \int_0^T dt & \{\vec{F}(t) \cdot [\vec{F}(t+\tau_1)\times\vec{F}(t+\tau_2)]\} \nonumber \\
&\cdot [\vec{F}(t+\tau_3)\cdot\vec{F}(t+\tau_4)]
\end{align}
and, in general,
\begin{align}
\label{eq_Hn}
H^{(n)}(\tau_1,\tau_2,...,\tau_{n-1}) = \int_0^T dt & \{\vec{F}(t) \cdot [\vec{F}(t+\tau_1)\times\vec{F}(t+\tau_2)]\} \nonumber \\
&... \, [\vec{F}(t+\tau_{n-2})\cdot\vec{F}(t+\tau_{n-1})]
\end{align}
where $n$ is an odd number.
For the field in Fig. \ref{fig_CPLvsSCL}b, the lowest-order non-zero chiral correlation function is $H^{(5)}$.
Therefore, the lowest order enantio-sensitive response of isotropic chiral matter to this light is of the fifth order.
In the frequency domain \cite{Ayuso2019NatPhot}
\begin{equation}\label{eq_h5}
h^{(5)}(-2\omega,-\omega,\omega,\omega,\omega) = \vec{F}^{*}_{2\omega} \cdot [\vec{F}^{*}_{\omega}\times\vec{F}_{\omega}] [\vec{F}_{\omega}\cdot\vec{F}_{\omega}]
\end{equation}
describes and quantifies the lowest-order enantio-sensitive response of isotropic chiral media to this light. Here, again, the enantio-sensitivity arises from the interference of two pathways.
The first, achiral, pathway is associated with the linear response at frequency $2\omega$, which leads to induced polarization at $2\omega$ along $z$.
In the second, chiral, pathway, the medium absorbs three $\omega$ photons from the major field component and emits one $\omega$ photon into the minor ellipticity component, also generating polarization at frequency $2\omega$ along $z$ (see  \cite{Ayuso2019NatPhot}), orthogonal
to the polarization plane of the $\omega$-field. 
The combination of up-- and down-arrows records the direction of rotation of the driving field.
The second pathway exists only in chiral media, and the induced polarization is out of phase in media of opposite handedness.
Interference between these two pathways enables enantio-sensitive absorption and emission at the frequency $2\omega$ and the possibility of achieving enantio-sensitive populations of excited electronic states.
The enantio-sensitive contributions to these observables can be written as a product of two pseudoscalars \cite{Ayuso2019NatPhot}: $h^{(5)}$ characterizing the field's handedness and a molecular pseudoscalar involving first- and fourth-order susceptibilities.$\blacksquare$

\textbf{Locally chiral light vs globally chiral light} While locally chiral light and the concept of local chirality  have been introduced \cite{Ayuso2019NatPhot} very recently, the second electric-dipole revolution started 
almost two decades ago.
Using quantum control strategies, Gerbasi, Brumer, Saphiro and co-workers proposed a two-step optical scheme for enantio-purification of randomly oriented mixtures of opposite enantiomers, which works in the electric-dipole approximation \cite{Gerbasi2004JCP}.
In the first step, a combination of three laser pulses with mutually orthogonal linear polarizations was used to selectively excite one of the two enantiomers to a selected vibrational state.
In the second step, the photo-excited molecules were forced to flip handedness by a sequence of two linearly polarized pulses.
Their simulations predicted $95\%$ of enantio-purity when starting from a racemic mixture of dimethylallene \cite{Gerbasi2004JCP}.
By controlling the relative phases between the laser fields in the first step, they were able to control whether the left-handed molecules were turned into right-handed or vice versa. This is probably the earliest example of application of locally chiral light to enantio-manipulation of molecules. The relative phases between the colors fully control the shape of light's Lissajous figure and its handedness,  controlling the outcome of the interference in Fig\ref{fig_ThreeLevelH3}.

Yet, there is an important caveat to this scheme. Locally chiral fields carrying three orthogonally polarized colours can be realized in the overlap region of two (or more) laser beams that propagate in different directions.
However, the phase delay between the non-collinear beams, i.e. the relative time at which their wavefronts reach a specific point in space, is space-dependent.
As a result, the handendess of the generated locally chiral field changes periodically in space, and so does the enantio-sensitive response of chiral matter.

For example,  for the laser parameters proposed in  \cite{Gerbasi2004JCP}, considering cross-propagating beams, the field's handedness would change in space with periodicities on the order of a few micrometers.
The application of such a field to a racemic mixture of isotropically distributed left- and right-handed molecules would create a non-homogeneous distribution of left- and right-handed molecules, which would be periodically distributed in space.
This structured distribution would be, on average, still racemic, unless using extremely tight laser focusing or thin media.

This problem is alleviated in the case of  the longer wavelengths associated with the microwave radiation, which leads to significantly wider spatial regions where the field maintains its local handedness, on the order of a few tens of centimeters. Eibenberger, Doyle and Patterson pioneered enantio-selective population of rotational states using phase-controlled microwave fields \cite{Eibenberger2017PRL} together with 
 Schnell and co-workers \cite{Perez2017}, who demonstrated an alternative implementation.

Unfortunately, if one tries to apply an equivalent scheme to achieve enantio-sensitive populations of \emph{electronic} states, one has to face the above problem:
the field's local handedness changes rapidly in space, destroying the enantio-sensitivity in the total (global) integrated response of the macroscopic medium.
To translate the huge enantio-sensitivity enabled by locally chiral fields to the macroscopic response of the medium, at the level of total signal intensities, the field also needs to be \emph{globally} chiral.

\textsf{\noindent\textbf{Concept 3. }
A field is \textbf{\emph{globally chiral}} if and only if its global (macroscopic) structure is chiral.}
\smallskip

While this definition is somewhat redundant when applied to standard circularly polarized light, which is either left- or right-handed everywhere in space,
it becomes particularly relevant in the characterization of the macroscopic response of isotropic chiral matter to locally chiral light, whose handedness can be spatially structured.
Such \emph{synthetic} chiral light is \textbf{globally chiral} if its handedness, characterized by the $n^{\text{th}}$-order chiral correlation function, survives integration in space, i.e. if $\int h^{(n)}(\vec{r}) d\vec{r} \neq 0$.

In particular, if the field's handedness is maintained in space, i.e. if the phase of $h^{(n)}$ remains constant, one can achieve the highest possible degree of control over the enantio-sensitive response of chiral matter: quench it in one enantiomer while maximizing it in its mirror twin \cite{Ayuso2019NatPhot}.

\begin{figure}[h]
\centering
\includegraphics[width=\linewidth, keepaspectratio=true]{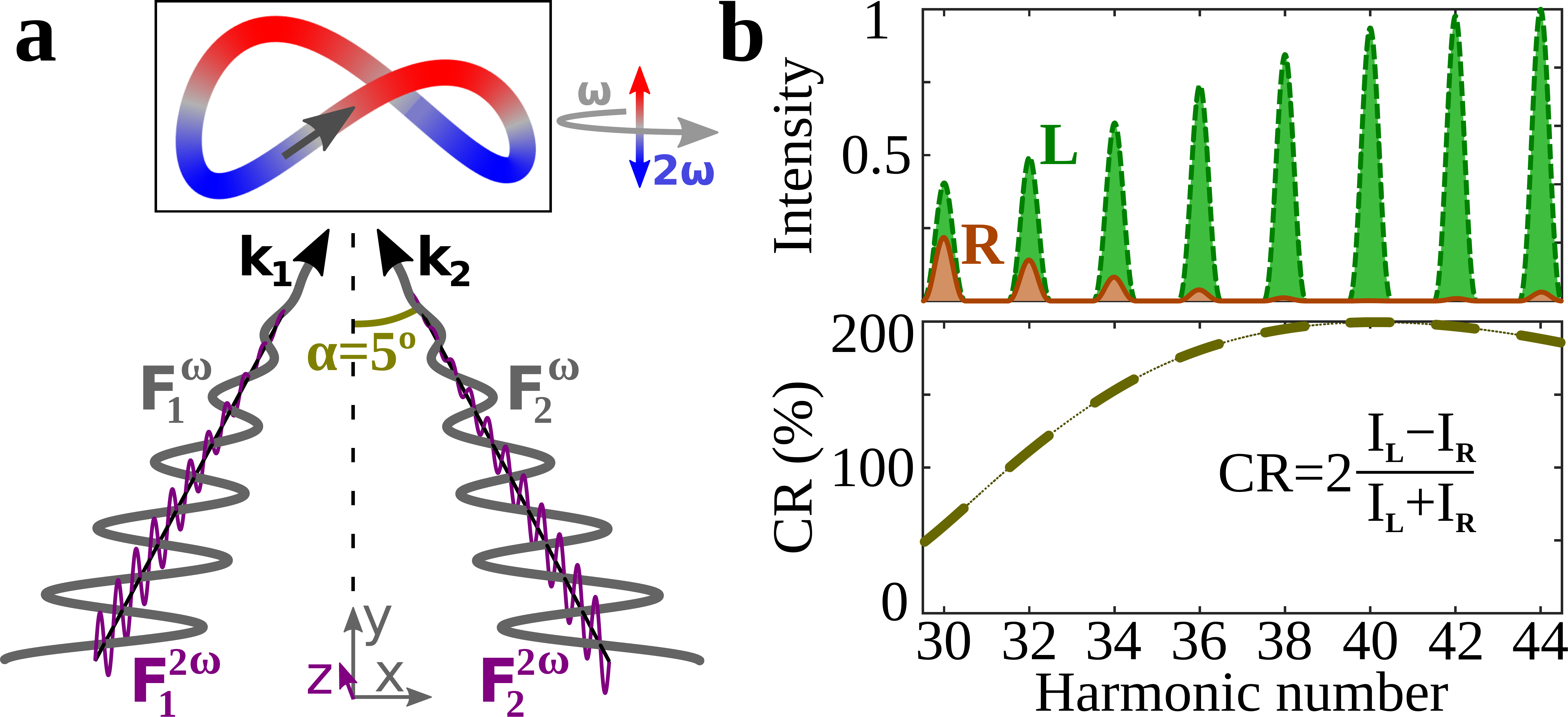}
\caption{\textbf{Locally and globally chiral light.}
\textbf{a}, Synthetic chiral light that is locally and globally chiral can be created with two non-collinear beams carrying cross polarized $\omega$ and $2\omega$ colours \cite{Ayuso2019NatPhot}.
In the overlap region, the total $\omega$ field is elliptical in the xy plane, the $2\omega$ field is z-polarized, generating the chiral Lissajous curve in the inset.
\textbf{b}, Even harmonic intensity emitted by randomly oriented left- and right-handed propylene oxide and the chiral response, see  \cite{Ayuso2019NatPhot} for details.
The field’s chirality, and thus the enantio-sensitive response of the medium, is fully controlled by the $\omega$,$2\omega$ phase delays in the two beams.
}
\label{fig_synthetic}
\end{figure}

The locally chiral field in Fig. \ref{fig_CPLvsSCL} can be created in a way that maintains the same handedness globally in space \cite{Ayuso2019NatPhot} using a non-collinear setup, see Fig. \ref{fig_synthetic}.
It contains two laser beams that propagate at a small angle, with each beam carrying two cross-polarized phase-locked colors: the fundamental and its second harmonic.
By controlling the $\omega$,$2\omega$ phase delays in the two beams, one achieves full control over the field’s local handedness globally in space.
This field enables complete discrimination between left- and right-handed randomly oriented chiral molecules via high harmonic generation spectroscopy \cite{Ayuso2019NatPhot}.

\section{New enantio-sensitive observables via structuring light's chirality}

Synthetic chiral light that is locally and globally chiral makes an extremely efficient chiral photonic reagent \cite{Ayuso2019NatPhot}.
Yet, the first dipole revolution taught us that we do not need to rely on the (global) handedness of light to detect the chirality of matter efficiently.
Can we apply these lessons to synthetic chiral light?
Can we measure strongly enantio-sensitive signals using light that is locally chiral ($h^{(n)}(\vec{r}) \neq 0$), but globally achiral ($\int h^{(n)}(\vec{r}) d\vec{r} = 0$)?
The answer is ``yes''. Applying the concepts from the first dipole revolution leads to new enantio-sensitive observables, which arise upon structuring light's local handedness.

\smallskip
\textsf{\noindent\textbf{Concept 4. \emph{Chirality-structured light}} is light whose handedness is non-trivially structured in space.}
\smallskip

The possibility of structuring the local properties of light in space \cite{Rubinsztein2016}, including both its intensity and phase \cite{Pisanty2019NatPhoton},  creates unique opportunities for imaging \cite{Hell2020} and manipulating \cite{Padgett2011} properties of matter.
Likewise, structuring light's chirality \cite{Tang2010,Patti2019,Li2019,Bradshaw2015,Cameron2014JPCA} could open new efficient routes for enantio-sensitive imaging and control of chiral matter.
With structuring of light performed locally, the 
control extends to the level of individual molecules.
One example of the new type of structured locally chiral light is the \emph{chirality polarized} light \cite{Ayuso2021NatComm}.

\begin{figure}[h]
\centering
\includegraphics[width=\linewidth, keepaspectratio=true]{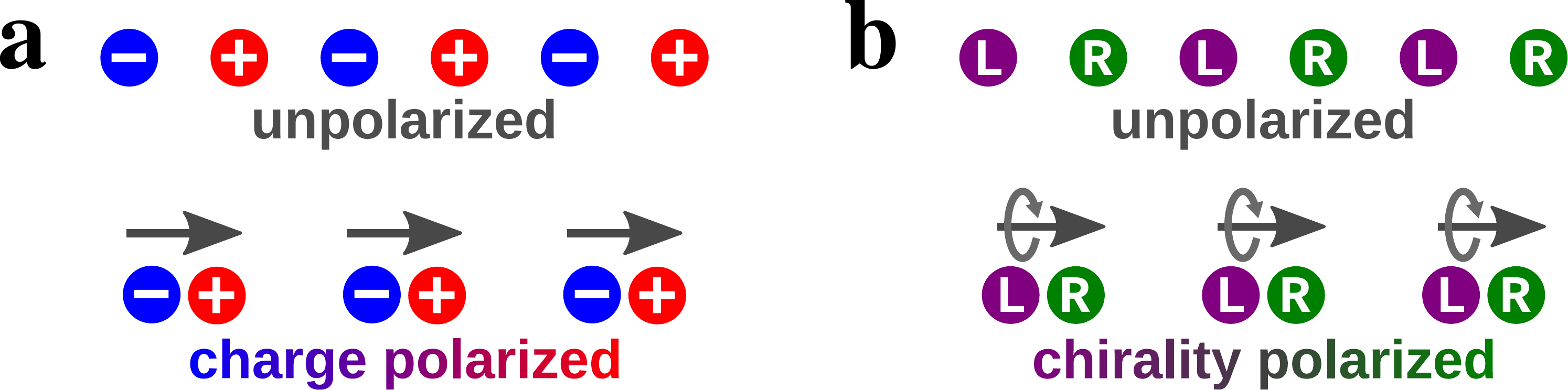}
\caption{\textbf{Polarization of charge versus polarization of chirality.}
\textbf{a}, 1D arrangement of charged units that is: (i) neutral and unpolarized, and (ii) neutral and polarized.
\textbf{b}, 1D arrangement of chiral units that is: (i) achiral (racemic) and unpolarized, and (ii) achiral and polarized.
}
\label{fig_polarization-concept}
\end{figure}
The concept of \emph{polarization of chirality} applies to both light and matter and is somewhat analogous to polarization of charge. A periodic distribution of alternating positive ($+q$) and negative ($-q$) charges in one dimension is unpolarized if the particles are uniformly distributed
and polarized 
if this distribution is periodically modified (Fig. \ref{fig_polarization-concept}a). 
Likewise, a periodic distribution of chiral units of alternating handedness can have polarization of chirality if the units are not uniformly distributed, see Fig. \ref{fig_polarization-concept}b.
Here, we find \emph{dipoles of chirality} $\vec{d}_c=h\vec{r}_0$, where $\vec{r}_0$ is the vector connecting two nearby chiral units and $h=h_R=-h_L$ is the handedness of one chiral unit.
Note that, regardless the value of $\vec{d}_c$, the medium is racemic and achiral, just like the medium of alternating negative and positive charges is neutral.

\begin{figure}[h]
\centering
\includegraphics[width=\linewidth, keepaspectratio=true]{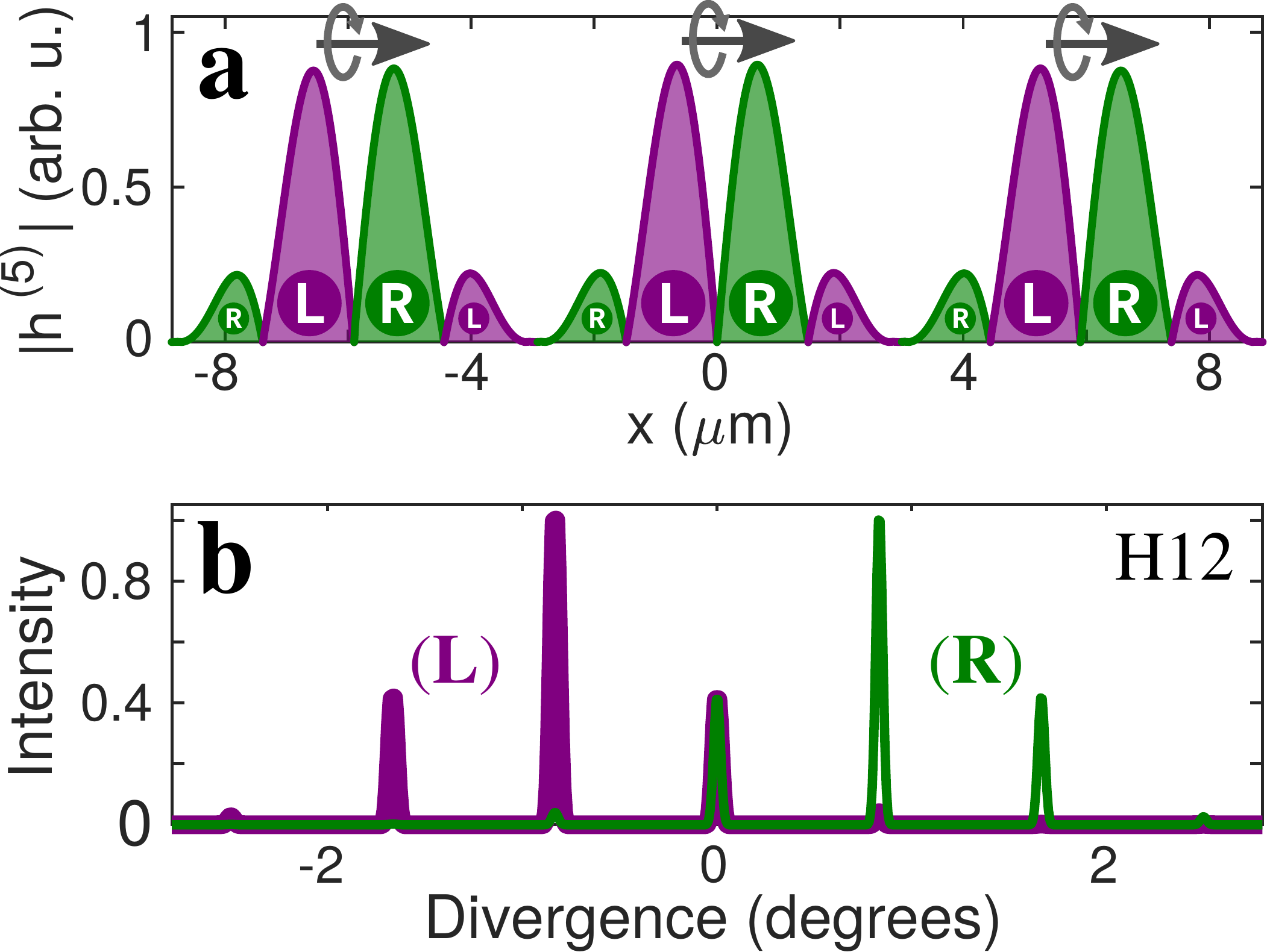}
\caption{\textbf{Chirality polarized light}
can be created using the setup of Fig. \ref{fig_synthetic}a, but adjusting the $\omega$,$2\omega$ phase delays so the field’s handedness, characterized by its 5th-order chiral correlation function $h^{(5)}$, is not maintained globally in space, in contrast to  \cite{Ayuso2019NatPhot}.
Here, it creates a periodic structure of dipoles of chirality.
\textbf{a}, h(5), its phase (i.e. the field’s handedness) is encoded in the colours. The arrows indicate the direction polarization of chirality, which is imprinted in the nonlinear response of chiral matter.
\textbf{b}, 12th-harmonic emission from randomly oriented fenchone, see  \cite{Ayuso2021NatComm} for $\lambda=1030$nm, $F_{\omega}^{(1)}=F_{\omega}^{(2)}=0.015$a.u. and $F_{2\omega}=F_{\omega}/10$.}
\label{fig_polarization}
\end{figure}

This concept can immediately be applied to the synthetic chiral light
using the same non-collinear setup as in Fig. \ref{fig_synthetic}a. Now, however, we can 
control the $\omega,2\omega$ delays in each beam so that the field's handedness, characterized by its fifth-order chiral correlation function, is not maintained globally in space
as in Ref. \cite{Ayuso2019NatPhot} but generates the ``dimers'' of alternating handedness: dipoles of chirality, see Fig. \ref{fig_polarization}a, polarized along x.

This periodic, racemic space-time structure will 
interact differently with chiral matter of opposite handedness by generating a periodic near-field pattern
of the enantio-sensitive nonlinear-optical local response. 
The mapping of this near-field nonlinear-optical response into the far-field image will be sensitive to 
the chirality dipole of light.

Fig. \ref{fig_polarization}b shows the harmonic intensity emitted from randomly oriented fenchone as a function of the emission angle.
The total (angle-integrated) signal is the same for left- and right-handed molecules, as the overall field is achiral.
However, the emission direction is highly enantio-sensitive: while the left-handed molecules emit light to the left (towards negative angles), the right-handed molecules emit light to the right (positive angles).

By looking into a specific emission direction, a chiral observer will allow us to distinguish between left- and right-handed molecules.
Note that neither the light field, nor the detector are chiral on their own.
However, by selecting a specific emission direction in the far field, the measurement setup, created by the (achiral) light field and the detector, becomes chiral.

\section{Towards efficient control, imaging  and manipulation of chiral molecules with light }

We now outline some of the new opportunities offered by substituting concepts and tools of enantio-sensitive molecular imaging and control, which use 
linear response and require the magnetic field component of the light wave, by
concepts and tools that rely on  non-linear response and
do not require the magnetic field component. In terms of imaging, in this section we address the opportunities for photon-based spectroscopies.
The next section will focus on new opportunities arising in photoelectron spectroscopy.

\subsection{Enantio-sensitive and molecular specific spectroscopy with synthetic chiral light}

In this subsection we describe the fundamental origins of 
enantio-- and molecular sensitivity in the interaction of
synthetic chiral light with chiral matter. Possible applications
of these ideas to enantio-separation are discussed in the subsequent subsection. 

\subsubsection{Enantio-sensitive molecular markers}
The enantio-sensitive response of chiral matter to synthetic chiral light relies on the coherent interplay between the two contributions to light-induced polarization,  achiral and chiral, the latter having opposite phase in media of opposite handedness.
One can control the amplitude and the relative phase of these two contributions by controlling the handedness of the light field, so that they interfere constructively in one enantiomer and destructively in its mirror twin, maximizing the enantio-sensitivity.

The next step is to develop spectroscopy, which would provide access to molecule-specific information recorded in such interference, such as the relative amplitude and phase between the chiral and the achiral non-linear responses. The latter 
should be naturally sensitive to the type of the molecule and its conformation  \cite{Ayuso2021ArXiv}. 
As a result, the handedness of the light field that maximizes enantio-sensitivity of the optical response should also be molecule-specific, opening new opportunities for efficient molecular recognition and enabling the design of \emph{molecular markers}: molecule-specific ``fingerprints'' of chiral molecules, based on the relative amplitude and phase of chiral and achiral non-linear responses.

For example, the intensity of the enantio-sensitive emission  at frequency $2\omega$ driven by the locally chiral field (Fig. \ref{fig_CPLvsSCL}) is
\begin{equation}\label{eq_interference_h5}
I = I_{ach} + I_{ch} + a\cos(\phi_M+\phi_{\omega,2\omega})
\end{equation}
where $I_{ach}$ and $I_{ch}$ are the intensities associated with the achiral and chiral pathways, respectively,
$\phi_M=\arg(\chi^{(4)}-\chi^{(1)})$ is the molecular phase that depends on the first- and fourth-order susceptibilities,
$\phi_{\omega,2\omega}$ is the 2-colour phase delay,
and $a \propto \sigma_M \sigma_L |\chi^{(1)}| |\chi^{(4)}| |h^{(5)}|$,
where $\sigma_M=\pm 1$ depends on the molecular handedness
and $\sigma_L=\pm 1$ keeps track of the sign of ellipticity of the light field.
Experimentally, one can vary $\phi_{\omega,2\omega}$ together with the amplitude of the $2\omega$-field component to find the laser parameters that maximize chiral dichroism.
These optimal parameters should be molecule-specific and may enable efficient molecular recognition.
Note that the phase $\phi_{\omega,2\omega}$ and the amplitude of the $2\omega$ field component controls the light pseudoscalar, which couples to the respective molecular pseudoscalar in the observables. Since both the light and molecular pseudoscalars are different in different non-linearity orders  \cite{Ayuso2019NatPhot}, corresponding to different frequencies of the emitted light, frequency-resolved optimal light parameters offer an additional  dimension (i.e. frequency) to this type of enantio-sensitive, molecule-specific spectroscopy.

\subsubsection{Ultrafast optical rotation: multi-dimensional nonlinear spectroscopy with sub-cycle-controlled optical waveforms.}

The sensitivity of an enantio-sensitive non-linear response to the two-colour phase (see Eq. \ref{eq_interference_h5}) points to new opportunities in non-linear spectroscopy with sub-cycle-controlled optical waveforms.

One example are few-cycle pulses,  
where the temporal structure of the electric field vector $\textbf{E}(t)=\textbf{E}_0a(t)\cos(\omega t+\phi_{\text{CEP}})$ depends on the carrier-envelope phase (CEP) $\phi_{\text{CEP}}$. The sensitivity of the  electronic dynamics to the instantaneous value of the oscillating electric  field leads to strong
CEP dependence of the nonlinear response \cite{Paulus2001Nat,Baltuska2003Nat,Schiffrin2013Nat,Luu2015}.
For our purposes, the CEP acts as an additional spectroscopic parameter somewhat equivalent to the two-colour phase delay in long laser pulses, which carry two well-defined frequency components such as $\omega$ and $2\omega$. In fact, changing the relative phase between these two colors shapes individual oscillations of the total laser field on the sub-cycle scale. 
Along this route, one can use intense, linearly polarized light pulses to drive the nonlinear analogue of optical rotation, now driven by purely electric-dipole interactions \cite{Ayuso2021Optica}.
Importantly, such pulses should be confined both in space and in time. 

Confinement in space can be achieved by tightly focusing the beam into a medium of randomly oriented chiral molecules, as shown in Fig, \ref{fig_One-color-Setup}b. Thanks to tight focusing, the field acquires ellipticity in the direction of light propagation, i.e.  ``forward ellipticity''.
The chiral medium converts this forward ellipticity into
the enantio-sensitive response orthogonal to the propagation plane (the plane of the Figure.)

Confinement in time, arising in nearly single-cycle pulses  with controlled CEP, ensures that the pulse has
ultra-broad spectral bandwidth.
The latter enables the interference of the
odd-order achiral response in the polarization plane and the even-order chiral response orthogonal to it. 

As a result, the generated nonlinear polarization becomes elliptically polarized and enantio-sensitive: the ellipticity and the rotation angle of the non-linear optical response will have opposite signs in media of opposite handedness, signifying the non-linear optical activity (rotation)  \cite{Ayuso2021Optica}. The rotation angle is controlled by the CEP of the laser pulse.
In contrast to conventional optical activity, this nonlinear effect is driven by purely electric-dipole interactions
and leads to giant rotation angles (and ellipticities) already at the single-molecule level, enabling the possibility of highly efficient chiral discrimination in optically thin  media.
Since the CEP of the  pulse controls the enantio-sensitive response in this non-linear optical activity, the value
of the CEP that maximizes the enantio-sensitive response of a chiral molecule may constitute a molecular marker.
The enantio-sensitive  signal plotted vs frequency and the CEP phase presents a two-dimensional set of data, which may be perceived as a ``molecular QR-code'', sensitive
to both the molecule and possibly its conformer. While the uniqueness of such measure is yet to be proven, the respective investigation is interesting in its own right, as it may result in a new interesting path in ultrafast non-linear spectroscopy.   
Adding additional spectroscopic parameters to a few-cycle laser pulse, such as a frequency-dependent phase delay  \cite{Yudin2006PRL}, may increase the dimensionality and consequently the sensitivity of this approach, extending opportunities for ultrafast enantio-sensitive imaging and control in the electric-dipole approximation.

\subsection{Enantio-sensitive manipulation.}

Locally and globally chiral electric fields create  opportunities for 
excitation of only one of the two enantiomers of a chiral molecule to  rotational, vibrational or electronic states and open routes to achieving efficient  enantio-manipulation.
Enantio-sensitive excitations to \emph{rotational} states have  already been demonstrated using microwave fields with three orthogonally polarized components \cite{Eibenberger2017PRL,Shubert2016,Perez2017,leibscher2020complete,leibscher2019principles} including opportunities for its optimal control  \cite{leibscher2020complete,leibscher2019principles}. 

A recent experiment  \cite{Milner2019PRL} demonstrated an alternative path to enantio-sensitive control over rotational excitations. This  experiment realises the concept of chiral observer and combines the optical centrifuge  \cite{Karczmarek1999PRL,Villeneuve2000PRL,Yuan2011PNAS} with Coulomb explosion imaging.  The chiral experimental setup consists of an optical centrifuge, i.e. a linearly polarised laser field rotating (with acceleration) at the slow, rotational, time scale about the propagation direction.
The additional axis is provided by the Coulomb explosion imaging detector. The respective molecular pseudoscalar  should include the projection of the total angular momentum (transferred from the field to the molecule and recorded in its rotational excitation) on the detector axis, i.e. on the direction of the Coulomb explosion.

Alternatively, one can also project the total angular momentum supplied by the optical centrifuge on a different axis, e.g. the direction of an additional external static field.
This setup has recently been proposed by Yachmenev et al \cite{Yachmenev2019PRL}. The combination of strong infrared fields forming the optical centrifuge and static electric field polarized along the centrifuge rotational axis is an example of locally and globally chiral electric field arrangement. 

Synthetic chiral light in the \emph{optical} domain enables control over the \emph{electronic} degrees of freedom, and thus the possibility of exciting a selected molecular enantiomer to a desired electronic or vibronic state.
Enantio-sensitive coherent control over such chiral electronic clouds opens additional routes for enantio-selective manipulation, since the properties of the photo-excited molecules, e.g. their dipole moments or polarizabilities, can be substantially different from those in the ground state.

Synthetic chiral light may also allow one to measure strongly enantio-sensitive photo-absorption signals, with the ambitious goal of enhancing the enantio-sensitivity of standard photo-absorption circular dichroism by several orders of magnitude.
Along this route, introducing synthetic chiral light to attosecond transient absorption spectroscopy (ATAS) may 
allow one to induce and measure strongly enantio-sensitive dynamics with attosecond time resolution using photon-based measurements.

\subsection{Enantio-separation.}

Intense laser beams allow one to manipulate, accelerate, decelerate and trap particles \cite{Bethlem1999PRL,Meerakker2008NatPhys,Maher-McWilliams2012NatPhot}.
Enantio-sensitive optical potentials, which could be applied for selective manipulation, trapping and sorting of chiral particles with specific handedness have recently been  demonstrated in linear light-matter interaction regime  \cite{Hernandez2013,CanaguierDurand2013,Tkachenko2014,Tkachenko2014_2,Hayat2015,Rukhlenko2016,Ali2020,Patti2019,Pellegrini2019}. 
A promising route to efficient enantio-separation was proposed by Cameron and 
co-workers \cite{Cameron2014NJP,Cameron2014JPCA}, who suggested creating gratings of chiral light of alternating handedness, which could then send opposite molecular enantiomers in opposite directions.
They found strongly enantio-sensitive deflection angles in chiral molecules with strong magnetic dipoles, such as helicene, demonstrating the feasibility of the method.
However, the proposed optical setups are chiral only beyond the electric-dipole approximation, and thus the sensitivity of this method to smaller molecules and localized chiral structures such as due to an asymmetric carbon, could be limited by the weakness of the non-electric-dipole interactions.
Synthetic chiral light may allow one to overcome this limitation, enhancing these opportunities even further.

\section{Synthetic chiral matter: Imprinting chirality on achiral matter}

Synthetic chiral light can also be used to endow achiral matter with chiral properties.
Achiral matter excited with such locally chiral light becomes ``synthetic chiral matter''.  That is, one can aim to imprint chirality on atoms  \cite{ordonez_propensity_2019}, achiral molecules  \cite{owens2018climbing}, and solids, and read it out on ultrafast scales.  
In molecules, Owens et al. \cite{owens2018climbing} showed that a chiral arrangement of a DC field and an optical centrifuge  \cite{Karczmarek1999PRL, Villeneuve2000PRL} produces $\mathrm{PH}_3$ molecules that rotate around a P--H bond in a direction determined by the centrifuge, with the P--H bond oriented along the DC field.  Formally, in this case the chirality of the electric field is imprinted on the molecular rotational states, which 
can be detected using e.g. ESMW spectroscopy.

In the case of electronic states, relatively simple superpositions of angular momentum eigenstates suffice to form chiral electronic wave functions in atomic hydrogen  \cite{ordonez_propensity_2019}, i.e. \emph{synthetic chiral atoms}, which
are already oriented in space. 
These chiral atoms can also display PECD upon ionization with circularly polarized field  \cite{ordonez_propensity_2019,ordonez_propensity_2022,mayer_imprinting_2021,katsoulis_momentum_2021}. The experimental demonstration of PECD on chiral atoms will require efficient excitation schemes. Such schemes have recently been proposed in the first theoretical works on using locally chiral light imprinting chirality on atomic ensembles  \cite{mayer_imprinting_2021,katsoulis_momentum_2021},  supporting our general expectation
that non-perturbative interaction of locally chiral light with atoms will in general lead to  chiral superpositions of states.

Interesting opportunities may arise for imprinting chirality on atoms in optical lattices and on electrons in solids. Recently, the spatial symmetry of the light's Lissajous figure has been used to imprint topological properties onto a trivial two-dimensional hexagonal material  \cite{jimenez-galan_lightwave_2020}. Analogously,
the longitudinal components emerging in tightly focused pulses can be used to make locally chiral light, which
can break the symmetry of a cubic lattice, by introducing ``forward-backward'' asymmetry in the field-modified hopping coefficients and thus turning the cubic lattice into a chiral object. Chiral crystals have been shown to possess interesting topological properties  \cite{chang_topological_2018}. Therefore, imprinting chirality on lattices might offer a way to induce light-driven ultrafast topology.

Looking broadly, the outcome of the interaction between such 
synthetic chiral matter and either `natural' matter or light (chiral or not) has not been explored and raises many interesting questions. How and to what extent will the chirality of synthetic chiral matter influence the outcome of its interaction with `natural' chiral matter or light? 
Is synthetic chiral matter useful for enantio-selective chemical synthesis? Can we design efficient catalysts for asymmetric synthesis based on synthetic chiral matter? Is it possible to implement ultra sensitive chirality detectors based on 
chiral Rydberg atoms? How will many-body effects influence the dynamics of chiral Rydberg atoms trapped in optical lattices? And can we perform quantum simulations of phenomena unique to chiral crystals \cite{sanchez_topological_2019} using chiral Rydberg atoms in optical lattices?

\section{Opportunities for imaging and control of chiral molecules via photoionization}
Photoelectrons are extremely sensitive structural probes of matter in gas and condensed phase. In molecules, electron scattering on the nuclei during photoionization provides information about their spatial arrangements. This information is recorded in the key photoionization observable: the angular and energy resolved photoelectron distribution; angle-resolved photo-electron spectroscopy can be viewed as ``diffraction from within.''

PECD and PXCD are two important milestones in understanding chiral electronic and vibronic response of molecules to light. PECD in photoionization from a stationary state is generated by chiral continuum currents, while PXCD in photoexcitation is generated by chiral bound currents. Probing PXCD via photoionization with a circularly polarized field would inevitably ``mix'' these two types of chiral dynamics and such time-resolved probe could be understood in terms of ``synchronisation'' of chiral bound and continuum currents. Controlling the interplay of pump and probe pulses can be interpreted in terms of controlling this synchronisation. Thus, formulating  standard photoionization observables in terms of the currents and fields they generate  \cite{ordonez_molecular_2020} may provide a helpful physical picture and a language for communicating  the observations. 

Since standard photoionization observables are defined in momentum space, the continuum electron currents and the fields they generate should also be defined in momentum space  \cite{ordonez_molecular_2020}.
Such perspective is customary in condensed matter physics, where one analyzes the $\vec{k}$ -dependent fields generated by electrons photo-excited from a valence to a conduction band. Electron scattering on the lattice sites may lead to  swirling electron trajectories. These, in turn, generate a geometric magnetic field, known as Berry curvature. The geometric magnetic field in solids plays an important role in understanding the electronic response, especially in materials with non-trivial topological properties.

Can the perspective offered by analysing the electronic response in molecular photoionization via a geometric magnetic field generated by swirling electron photoionization current be fruitful, especially for chiral molecules? Could it allow us to uncover new enantio-sensitive  observables in photoionization? What can we learn from the topology of field lines?
How does this field reflect information about molecular dynamics prior to ionization?
Interestingly, the standard photoionization observables formulated in terms of local, i.e. $\vec{k}$-dependent currents and the fields they generate, provide some answers to these questions (see Section 8.4 below).

As an extremely sensitive probe of molecular chirality,  PECD,  has already given rise to a family of time-resolved methods which use various pump-probe setups and detect angularly resolved photoelectron spectra. All these methods involve two- or higher order multiphoton processes. An important player in such setups is the polarization of pump and probe pulses. For example, two-photon ionization of a chiral molecule can be performed in many different ways:  linearly polarized pump - circularly polarized probe (TD PECD  \cite{Comby2016JPCL}), circularly polarized pump - linearly polarized probe (PXECD  \cite{Beaulieu2018PXCD}),  circularly polarized pump - circularly polarized probe  co-rotating (coherent control of PECD  \cite{Goetz2019PRL,ordonez_molecular_2020}) or counter-rotating to each other.

Interestingly, chiral properties can also be probed using linearly polarized pump and probe pulses, but conditions apply.
What are these conditions? What is the difference in photoionization observables detected using these different schemes?

In this section, we first provide the perspective on such two-photon measurements, which demonstrates that each of these schemes addresses different molecular properties, because it relies on different molecular pseudoscalars and therefore exposes a different and independent aspect of molecular chirality.  Next, we discuss the physical picture behind these schemes and  perspectives for imaging and control of chiral molecular dynamics inspired by these schemes. Finally, we discuss the geometric magnetic field and the physical origin of PECD using the language of currents and fields generated by  electrons during photoionization. At this point we would also address the X-ray community, which identifies imaging of molecular currents as an important milestone.

\subsection{Chiral molecular fingerprints in the two-photon angular resolved photoionization}
In these subsection we are dealing with methods of chiral detection relying on the principle of the chiral observer, just like the  PECD and PXCD methods considered in Section 3. However, there is an important difference.
While PECD and PXCD are one-photon processes and involve two light field vectors that form a single light pseudovector $\vec{L}$ (see Eq. \ref{eq:v=gL}), PEXCD is a two-photon process. In two-photon and, in general,  multiphoton processes more than two light vectors 
may be relevant. Therefore,  different pseudovectors $\{\vec{L}_1, \vec{L}_2, \dots\}$ can be formed. This gives rise to a \emph{hierarchy of chiral measures} (see Section \ref{sec:hierarchy}). Analogously, more than three molecular vectors may be relevant, and thus several different molecular pseudoscalars $\{g_1, g_2, \dots\}$ can be formed. The availability of multiple light pseudovectors and multiple molecular pseudoscalars is reflected in a generalization of Eq. (\ref{eq:v=gL}) according to
 \begin{equation}
    \vec{v}=\sum_i g_i \vec{L}_i \ \ .
     \label{eq:v=gL_general}
 \end{equation}
That is, instead of a single product between a molecular pseudoscalar and a light pseudovector, the enantio-sensitive vectorial observable $\vec{v}$ is given by a sum of several  such products, involving all possible combinations of all available vectors.
Rule 7 in Section \ref{sec:hierarchy} provides the most general expression and points to its formal origins.
 
\begin{figure}[h]
\centering
\includegraphics[width=\linewidth, keepaspectratio=true]{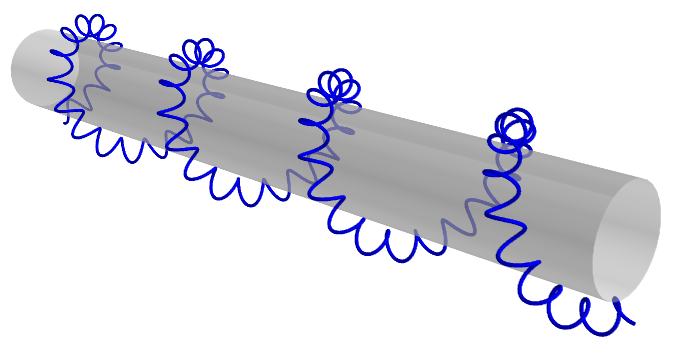}
\caption{An object displaying compound chirality in the form of two independent handedness: a helix made of a more tightly
bound helix.}
\label{fig:helix_of_a_helix}
\end{figure}

The array of different molecular pseudoscalars simply reflects the fact that the description of a complex chiral object requires the specification of more than one pseudoscalar. For example, while a simple helix has just one helicity, a compound helix where a loosely wound helix is formed from a tightly wound helix requires the specification of two independent helicities: one for the loose helix and one for the tight helix (see Fig. \ref{fig:helix_of_a_helix}). In this sense, the coupling between molecular pseudoscalars and light pseudovectors in Eq. $(\ref{eq:v=gL_general})$ reveals how each of the `helicities' of a chiral molecule (a rather complex `compound helix') couples to the light field. (Similar situation with "nested" chiral structures can occur due to the presence of several chiral centers with different handedness in a chiral molecule).
 
Two photon pump-probe process are a perfect example illustrating Eq. (\ref{eq:v=gL_general}).  Suppose that the pump $\vec{E}_1$ induces a transition from the ground state $|g\rangle$ into  a bound excited state $|j\rangle$ and the probe $\vec{E}_2$ induces a transition from the state $|j\rangle$ into the continuum $|\vec{k}\rangle$. For pulses and time delays short compared to the rotational dynamics, the photoelectron current is given by \cite{Ordonez2018generalized}
\begin{equation}
\vec{J} = \sum_{i=0}^{6} g_i \vec{L}_i,
\label{eq:j_pumpprobe}
\end{equation}
where the characteristic molecular pseudoscalars $g_i$ are:
\begin{align}
g_0 =&\frac{1}{30}\left|\vec{d}_{j,g}\right|^{2} \int\mathrm{d}\Omega_{k}\left[\left(\vec{d}_{\vec{k},j}^{*}\times\vec{d}_{\vec{k},j}\right)\cdot\vec{k}\right], \label{eq:g0}\\
g_1 =& \frac{1}{30} \int\mathrm{d}\Omega_{k}\left[\left(\vec{d}_{\vec{k},j}^{*}\times\vec{d}_{\vec{k},j}\right)\cdot\vec{d}_{g,j}\right]\left(\vec{d}_{j,g}\cdot\vec{k}\right),\\
g_2 =& \frac{1}{30} \int\mathrm{d}\Omega_{k}\left[\left(\vec{d}_{g,j}\times\vec{d}_{\vec{k},j}^{*}\right)\cdot\vec{k}\right]\left(\vec{d}_{j,g}\cdot\vec{d}_{\vec{k},j}\right),\label{eq:g_2}\\
g_3 =& \frac{1}{30} \int\mathrm{d}\Omega_{k}\left[\left(\vec{d}_{i,j}\times\vec{d}_{\vec{k},j}\right)\cdot\vec{k}\right]\left(\vec{d}_{j,g}\cdot\vec{d}_{\vec{k},j}^{*}\right),
\label{eq:g_3}
\end{align}
with 
\begin{equation}
g_4 = g_1^*,\quad g_5 = g_2^*,\quad g_6 = g_3^*.
\end{equation}
Importantly, these pseudoscalaras couple to different light pseudovectors:\begin{align}
\vec{L}_0 = \left|\vec{E}_1\right|^{2} \left(\vec{E}_2^{*}\times\vec{E}_2\right),\quad
&\vec{L}_1 = \left[\left(\vec{E}_2^{*}\times\vec{E}_2\right)\cdot\vec{E}_1^{*}\right]\vec{E}_1, \label{eq:L0,L1}\\
\vec{L}_2 = \left(\vec{E}_1\cdot\vec{E}_2\right)\left(\vec{E}_1^{*}\times\vec{E}_2^{*}\right),\quad
&\vec{L}_3 = \left(\vec{E}_1\cdot\vec{E}_2^{*}\right)\left(\vec{E}_1^{*}\times\vec{E}_2\right),\label{eq:L2,L3}
\end{align}
\begin{equation}
\vec{L}_4 = \vec{L}_1^*,\quad \vec{L}_5 = \vec{L}_2^*,\quad \vec{L}_6 = \vec{L}_3^*, \label{eq:L4,L5,L6}
\end{equation}
which define constrains on the observation of each molecular pseudoscalar $g_i$ in terms of polarization of pump and probe pulses.

For example,  the $g_0 \vec{L}_0$ term is associated with the PECD from the excited state $|j\rangle$ and requires only an elliptically polarized probe field. The pseudovector fomed by photoionization vectors $(\vec{d}_{\vec{k},j}^{*}\times\vec{d}_{\vec{k},j})$ is important in its own right, since it can be understood as a geometric magnetic field $\vec{B}(\vec{k})$ associated with photoionization of chiral molecules  \cite{ordonez_propensity_2019} (see subsection \ref{subsec:geometric_magnetism}). Note that $g_0 $ and $g_1 $ encode different components of $\vec{B}(\vec{k})$: $g_0$ encodes the radial field component and  $g_1$ encodes the field component along the direction of the bound dipole connecting the ground and excited states. 

Accessing $g_1 \vec{L}_1$ requires a different and more sophisticated arrangement of pump and probe polarizations. Indeed, $\vec{L}_1$ is non-zero only when $\vec{E}_2$ is elliptically polarized and in addition $\vec{E}_1$ has a component perpendicular to the plane defined by $\vec{E}_2$. In contrast, $\vec{L}_2$ and $\vec{L}_3$ are non-zero even if both the pump and the probe are linearly polarized, as long as they are neither parallel nor perpendicular to each other, suggesting a connection to some recent works using  such arrangement of pump and probe pulses to  excite rotational states of chiral molecule and induce 
enantio-sensitive molecular orientation  \cite{Yachmenev2016PRL,Gershnabel2018PRL,Tutunnikov2019PRA}. Note also that while $\vec{L}_2$ vanishes for co-rotating circularly polarized pump and probe, $\vec{L}_3$ vanishes for counter-rotating pump and probe.
These simple rules together with the [Eqs. (\ref{eq:j_pumpprobe})-(\ref{eq:L4,L5,L6})]  specifying the coupling between molecular pseudoscalars and field pseudovectors can be combined to determine the values of each molecular pseudoscalar $g_i$. 

In the case where the final state $|\vec{k}\rangle$ can be reached via two different intermediate states, additional molecular pseudoscalars  and field pseudovectors (see Ref.  \cite{Ordonez2018generalized}) resulting from the two-path interference contribute to the generation of the photoelectron current. Importantly, they also can be expressed  \cite{Ordonez2018generalized} in  compact form similar to Eqs. (\ref{eq:g0})-(\ref{eq:L4,L5,L6}).
 Unlike the direct terms, these terms oscillate with the time delay between pump and probe and record the dynamics  excited by the pump. One example of such dynamics are helical currents excited in bound states by ultrashort circualry polarized pulses (PXCD). Below we provide a perspective on their detection.
    
 \subsection{Nonlinear photoionization probes of molecular chirality}
 
In the previous section we gave the formal description of the two-photon pump-probe setups for probing chiral molecular structure and dynamics. Here we discuss the underlying physical picture describing two-photon photoionization  as an interplay of chiral bound and continuum currents. This interplay can be disentangled in PXECD, which induces photoionization using linearly polarized light to decrease the influence of continuum currents.
 
\textbf{Probing helical currents in bound states (PXECD) via photoionization.}
Helical currents excited in chiral molecules by short circularly polarized pulses (PXCD, see Section 3) can be probed by photoionization. Since induced polarization oscillates out of phase in opposite enantiomers, the respective electron currents flow in opposite directions. Photoionization by linearly polarized short pulse can reveal this direction of the bound current: the liberated electron should typically continue to move in the same direction as it was moving in the bound states. Thus, two photoelectron detectors placed along the propagation direction (spin) 
of the pump pulse will show the forward-backward asymmetry: more forward electrons in one enantiomer and more backward electrons in the opposite enantiomer.

Interference of two photoionization signals originating from the two excited PXCD states (see Fig. \ref{fig_PXCDvsPECD}) records the coherence between these states and also its time evolution as a function of the delay between the exciting (pump) and photoionizing (probe) pulses. Since both pulses are short, interference of photoionization signals coming from the two intermediate states excited by the pump may cover a considerable range of photoelectron energies. In this range, one can observe that the photoelectron current oscillates with the time-delay between the two pulses and has opposite directions for opposite enantiomers, or for opposite spins of the pump pulse. That is, thanks to the molecular chirality, the spin of the pump photon can be read out from the photoelectron angular distribution even though the ionizing step is carried out with linearly polarized light. 

Indeed, one can show \cite{Beaulieu2018PXCD} that the photoelectron current is proportional to two terms:
\begin{eqnarray}
 \label{PXECD}
 J_z^{\mathrm{PXECD}}(k)=\sigma[(\vec{d}_{01}\times \vec{d}_{02})\cdot\vec{D}_{12}^{r}(k)]\sin(\Delta E_{21}\tau)-\nonumber\\
 \sigma[(\vec{d}_{01}\times \vec{d}_{02})\cdot\vec{D}_{12}^{i}(k)]\cos(\Delta E_{21}\tau), 
 \end{eqnarray}
where $J_x^{\mathrm{PXECD}}(k)=J_y^{\mathrm{PXECD}}(k)=0$,
$\hat{z}$ and $\sigma$ are the propagation direction and spin of the pump pulse, respectively,
$\tau$ is the pump–probe delay, $k$ is the photoelectron momentum, $\Delta E_{21}$ is the energy difference between the excited states, $\vec{d}_{01}$ and $\vec{d}_{02}$ are the transition dipoles from the ground state to the excited states, $\vec{D}_{12}(k)=\vec{D}^r_{12}(k)+i\vec{D}^i_{12}(k)$ is a complex  Raman-type photoionization vector  that connects excited bound states via the common continuum. Since $\vec{D}_{12}(k)$ encodes coherence between the excited states, it plays the role of $\vec{d}_{12}$ in Eqs. (\ref{eq:gL_PXCD})-(\ref{eq:L_PXCD}) for PXCD. 
The two triple products in Eq. (\ref{PXECD}) are the molecular pseudoscalars characterizing the enantio-sensitive signal in PEXCD. 
The photoionization vector $\vec{D}_{12}$ in Eq. (\ref{PXECD}) is given by
 \begin{eqnarray}
\vec{D}_{12}(\vec{k}) \equiv -4\left(\vec{D}_{1}\cdot\vec{D}_{2}^{*}\right)\vec{k}
+\left(\vec{D}_{2}^{*}\cdot\vec{k}\right) \vec{D}_{1}
+\left(\vec{D}_{1}\cdot\vec{k}\right)
\vec{D}_{2}^{*}.
\label{D12_veck_simplified}
\end{eqnarray}

Note that this general expression shows that every available vector $(\vec{D}_{1},\vec{D}_{2}^{*},\vec{k})$ can be used to ``complete'' the triple product in Eq. (\ref{PXECD}). Physically, the appearance of the second and the third term in Eq. (\ref{D12_veck_simplified}) is due to partial alignment of molecules by the pump pulse. To validate this statement it is sufficient to consider an isotropic probe pulse, which cannot be sensitive to the initial alignment by the pump. One can show (see SI of Ref.  \cite{Beaulieu2018PXCD}) that only the first term in Eq. (\ref{D12_veck_simplified}) survives in this case.

How accurately can the continuum current image the current in 
the bound states? The difficulties here are similar to those encountered in the so-called tomographic imaging of molecular orbitals in High Harmonic Spectroscopy  \cite{itatani_tomographic_2004}. Namely, the bound-continuum mapping is only exact in the case of the plane wave continuum: in this case,  the total photoelectron current $\vec{J}^{\mathrm{PXECD}}_{\mathrm{tot}}$, integrated over all photoelectron energies $\vec{J}^{\mathrm{PXECD}}_{\mathrm{tot}}\equiv \int \vec{J}^{\mathrm{PXECD}}_{\mathrm{PW}}(k)dk$ is indeed proportional to the bound current $\vec{J}^{\mathrm{PXECD}}_{\mathrm{tot}}\equiv \vec{J}^{\mathrm{PXCD}}$,  since one can show that $-(1/2)\int \vec{D}^{i,PW}_{12}(k)dk\equiv\Delta E_{12}\vec{d}_{12}$. However, the structure of the continuum in chiral molecules is a lot more complex than simple plane waves.  
Perhaps, the connection between the
bound and continuum currents can be established by introducing an additional unknown function $f(k)$, such that $-(1/2)\int \vec {D}^{i,\mathrm{PW}}_{12}(k)f(k)dk\equiv \Delta E_{12}\vec{d}_{12}$. This function will then have to be reconstructed together with the bound current, e.g. iteratively, starting with $f(k)=1$ for the plane wave continuum, similar to the efforts in tomographic reconstruction of molecular orbitals in High Harmonic Spectroscopy  \cite{vozzi_generalized_2011}.

The first  PEXCD images \cite{Beaulieu2018PXCD} were recorded via excitation of Rydberg bands in fenchone and camphore molecules using a circularly polarized femtosecond pump pulse carried at 201 nm (with 80 meV $1/e^2$ bandwidth) and probing it using a time-delayed, linearly polarized probe pulse carried at 405 nm (with 85 meV at $1/e^2$ bandwidth). Although the results demonstrate excitation of a chiral vibronic wave-packet in these molecules, detailed information about the specific nature of these dynamics requires further analysis. Such analysis could provide much desired insight into chiral molecular dynamics at femtosecond time-scales and presents one of the exciting future opportunities for this field.$\blacksquare$

\textbf{Two-color coherent control.} Another interesting aspect of chiral molecular dynamics may result from the interplay of bound and continuum chiral currents. Both currents are present if both pump and probe pulses are circularly polarized. Indeed,  the photon spin carried by the pump can be transferred to the current excited in bound states. At the same time, the photon spin carried by the probe pulse can lead to chiral continuum currents. This combination of pump and probe polarizations  has been recently explored by  Goetz et al  \cite{Goetz2019PRL} and used for two-photon coherent control  of the chiral photoelectron current associated with the coefficient $b_{1,0}$ and multipolar currents associated with the coefficient $b_{3,0}$.  Goetz et al  \cite{Goetz2019PRL} achieved very significant enhancement of enantio-sensitivity of photoionization observables by  optimizing the arrival of each frequency in the pump and probe pulses. Since the scheme involves the absorption of two circularly polarized photons,  the optimization could have been related to achieving the best synchronization between the bound and continuum currents.
Whether the control  has been  associated with such synchronisation remains to be seen, but exploring and exploiting the interplay of bound and continuum currents for enhancing chiral photoelectron signal may present  an interesting future direction.$\blacksquare$
   
\textbf{Time-dependent PECD} implies exciting molecular vibronic dynamics by a linearly polarized pump pulse and probing it with a circularly polarised pulse  \cite{Comby2016JPCL}. As we have seen from previous examples, excitation with a circularly polarized pump imprints the photon spin onto the bound states dynamics and excites helical currents in chiral molecules. At first glance one may think that a linearly polarized pump cannot excite chiral dynamics, because the spin is required to define the ``helicity'' of the excited bound current. Indeed, since a linearly polarized pump is a superposition of two counter-rotating circular pulses, we should expect the excitation of two PXCD currents of opposite handedness in a given enantiomer. However, the probe  -- a circularly polarized pulse, will break this symmetry between left and right helical PXCD currents. Indeed, two co-rotating photons and two counter-rotating photons produce different photoionization signals, because they correspond to different molecular pseudoscalars [see e.g. Eqs. (\ref{eq:g_2}) and (\ref{eq:g_3}) for $g_2$ and $g_3$ in case of a single intermediate state.  Eq. (35) of ref.  \cite{Ordonez2018generalized} generalizes this result for two intermediate states].  Thus, the time-dependent PECD is a probe highlighting the synchronisation of bound and continuum chiral currents pondered above. It presents a differential measure encoding two chiral currents of opposite handedness with different amplitudes. The results of time-dependent PECD experiments could be re-interpreted as the images of such interplay   \cite{Comby2016JPCL}. 

\subsection{Enantio-separation via photoionization}

The possibility of selectively exciting only one of the two enantiomers of a chiral molecule to an electronic state opens several routes for enantio-separation that we chart below.

One option is to  \textbf{use synthetic chiral light} \cite{Ayuso2019NatPhot} (See section \ref{sec:chiral_reagent}) to selectively excite one of the two enantiomers of a chiral molecule to a desired state, tuning the frequencies of the light field in (possibly multi-photon) resonance with a specific electronic or vibronic transition.
Next, the photo-excited molecules could be photoionized with a second laser pulse, yielding enantio-selective ionization.
These molecular ions with well-defined handedness could then be extracted with a static field.  

Another option is to \textbf{ use achiral light without photon spin} to achieve enantio-sensitive uniaxial orientation of chiral molecules on the electronic time-scale  \cite{ordonez_inducing_2021}.

It is often assumed that molecular orientation can only occur on rotational time-scales. However, in chiral molecules, this does not have to be the case. Our analysis  \cite{ordonez_inducing_2021} in the perturbative regime predicts that phase-locked, orthogonally polarized fields with frequencies $\omega$ and $3\omega$ can induce field-free permanent electronic dipoles in initially isotropic samples of chiral molecules via resonant electronic excitation. The dipole’s orientation is enantio-sensitive and it is controlled by the relative phase between $\omega$ and $3\omega$ fields, which determines the sub-cycle direction of rotation of the total electric field. In contrast to the photo-excited circular dichroism (PXCD)  \cite{Beaulieu2018PXCD}, here not only the excited electron but also the molecule correlated to the excitation acquires orientation. 

This effect is fundamentally multi-photon. In the frequency domain, the interference between the two pathways, $3\times \hbar\omega$ vs. $1\times3\hbar\omega$, is sensitive to the molecular orientation and handedness. This leads to orientation-dependent excitation and thus uniaxial orientation of the excited molecules, on the electronic excitation time scale. The orientation is perpendicular to the polarization plane and is reflected in the emergence of a field-free permanent dipole. 
 
This fundamental phenomenon points to interesting opportunities for creating enantio-sensitive permanent dipoles via orientation-dependent excitation of Rydberg states
upon resonance-enhanced multiphoton ionization (REMPI). The non-perturbative $\omega$ and $3\omega$ fields should also be explored, since in such regime hitting resonances does not necessarily require carefully tuning the light frequency to specific molecular transitions. Indeed, in the non-perturbative regime one can take advantage of light-induced energy shifts of excited states. These are known as Freeman resonances  \cite{freeman_above-threshold_1987} and are virtually inevitable at intensities $I\sim 10^{13}$ $\mathrm{W/cm}^2$ and above. They will also lead to orientation of molecular ions after orientation-selective resonantly enhanced multi-photon ionization. By selectively depleting randomly oriented neutrals, preferential orientation in the neutral ensemble is also created. After that a static field can be used to spatially separate opposite enantiomers.

\subsection{Geometric magnetism in chiral molecules}\label{subsec:geometric_magnetism}

The excitation of enantio-sensitive photoelectron currents (PECD) in the electric-dipole approximation can be linked to the concept of the geometrical magnetism introduced by M. Berry  \cite{berry1984quantal}.  One of its manifestations is the Berry curvature in solids, which enables a class of phenomena in condensed matter systems including anomalous electron velocity, the Hall effect, and related topological phenomena  \cite{resta_macroscopic_1994}.  A geometric magnetic field also appears in photoionizaton of chiral molecules by circularly polarized fields. 
This field arises due to ``curly'' or ``twisted'' polarization in vibronic states or due to ``curly'' or ``twisted'' currents. The ``twist'' originates from the chiral arrangements of the nuclei and does not vanish upon averaging over the random molecular orientations. The geometric magnetic field arising in chiral molecules underlies several classes of chiral photoionization observables. It is related to the so-called propensity field that we have introduced recently  \cite{ordonez_propensity_2019-2}.

\textbf{The propensity field in photoionization} involves the vector product of two conjugated photoionization dipoles, \cite{ordonez_propensity_2019-2}
\begin{equation}
  \vec{B}(\vec{k})=i[\vec{d}_{\vec{k},g}\times \vec{d}^*_{\vec{k},g}]=i\frac{[\vec{p}_{\vec{k},g}\times \vec{p}^*_{\vec{k},g}]}{(E_k-E_g)^2}.
\label{eq:fieldB}
\end{equation}
where $\vec{d}_{\vec{k},g}$ and $\vec{p}_{\vec{k},g}=i(E_k-E_g)\vec{d}_{\vec{k},g}$ are transition dipoles in the length and velocity gauges, respectively, and $E_k$ and $E_g$ are the energies of the photoelectron and of the ground state, respectively.
        
As usual for photoionization observables, the propensity field $\vec{B}(\vec{k})$ is a function of the photoelectron momentum $\vec{k}$. $\vec{B}(\vec{k})$ quantifies the absorption circular dichroism (CD) for a specific state $|\vec{k}\rangle$. The direction of $\vec{B}(\vec{k})$ indicates a preferred direction in the molecular frame: if a circularly polarized light propagates along this direction, it maximizes the CD for a transition from the ground state into a specific final state $|\vec{k}\rangle$.
Loosely speaking, the direction of $\vec{B}(\vec{k})$ defines the axis in the molecular frame along which the rotational symmetry of the molecule is broken to the highest extent, for a given final state $\vec{k}$. 

The magnitude of $\vec{B}(\vec{k})$ is proportional to the corresponding CD, i.e. it is proportional to the difference between the populations of the state $|\vec{k}\rangle$ obtained with left and right circularly polarized light propagating along $\vec{B}(\vec{k})$.
Indeed, if we denote the direction of $\vec{B}(\vec{k})$ by  $\hat{e}_B(\vec{k})\equiv\vec{B}(\vec{k})/{|B(\vec{k})|}$, we obtain \cite{ordonez_propensity_2019-2, ordonez_molecular_2020, ordonez_geometric_2021} 
\begin{equation}
  \vec{B}(\vec{k})\cdot \hat{e}_B=|\vec{d}^{+}_{\vec{k},g}|^2-|\vec{d}^{-}_{\vec{k},g}|^2, \hspace {0.5 cm} \hat{e}_B\equiv\frac{\vec{B}(\vec{k})}{|B(\vec{k})|},
\end{equation}
where $\vec{d}^{\pm}_{\vec{k},g}$ are the photoionization dipoles for ionization by left or right circularly polarized fields propagating along the direction specified by $\vec{B}(\vec{k})$.
Thus, the vector field $\vec{B}(\vec{k})$ provides the molecule-specific  $\vec{k}$-resolved map of maximal possible photoionization CD.

One can show that the propensity field is analogous to the Berry curvature in two band solids,  \cite{ordonez_propensity_2019-2, ordonez_geometric_2021} which is responsible for the circular photogalvanic effect in chiral solids  \cite{de_juan_quantized_2017} in the same way as the propensity field is responsible for PECD in gas phase chiral molecules. The chiral geometric magnetic field introduced in Ref.  \cite{ordonez_geometric_2021} provides a generalization of the propensity field.$\blacksquare$ 
        
\textbf{Geometric field in molecular photoionization \cite{ordonez_geometric_2021}} The propensity field is related to photoionization from a single state. In a more general case of photoionization from a superposition of two excited states $|j\rangle+e^{-i\phi_{ij}}|i\rangle$ ($\phi_{ij}=\omega_{ij}t$, with $\omega_{ij}\equiv\omega_i-\omega_j$ being the transition frequency between the states), the 
propensity field encodes the coherence between the excited states:
\begin{align}
\label{eq:B_ij}
\vec{B}_{ij}(\vec{k},\phi_{ij})=-\frac{1}{2}i\left[\vec{d}^{*}_{\vec{k}i}\times \vec{d}_{\vec{k}j}\right]e^{i\phi_{ij}} + \mathrm{c.c.} \nonumber \\
\equiv\vec{Q}_{ij}(\vec{k})\cos\phi_{ij}+\vec{P}_{ij}(\vec{k})\sin\phi_{ij},
\end{align}
where we  have introduced the displacement $\vec{Q}_{ij}(\vec{k})$ and current $\vec{P}_{ij}(\vec{k})$ quadratures: 
\begin{equation}
\vec{Q}_{ij}(\vec{k})\equiv -\Re\left\{i\left[\vec{d}^{*}_{\vec{k}i}\times \vec{d}_{\vec{k}j}\right]\right\},
\label{eq:Q_ij_definition}
\end{equation}
\begin{equation}
\vec{P}_{ij}(\vec{k})\equiv \Im\left\{i\left[\vec{d}^{*}_{\vec{k}i}\times \vec{d}_{\vec{k}j}\right]\right\}.
\label{eq:P_ij_definition}
\end{equation}
Eq. (\ref{eq:B_ij}) describes the geometric field oscillating at the frequency $\omega_{ij}$. For $i=j=g$, $\phi_{ij}=0$ and
Eq. (\ref{eq:B_ij})  reduces to Eq. (\ref{eq:fieldB}). For any number of states  Eqs. (\ref{eq:fieldB}) and (\ref{eq:B_ij}) can be generalized as
\begin{equation}
\label{eq:B_ij_sum}
\vec{B}(\vec{k},t)=\frac{1}{2}\sum_{i,j}\left\{i\left[\vec{d}_{\vec{k}j}\times \vec{d}^{*}_{\vec{k}i}\right]\right\}e^{i\phi_{ij}}.
\end{equation}
We call $\vec{B}(\vec{k},t)$ in Eq. (\ref{eq:B_ij_sum}) the \emph{geometric field in molecular photoionization}.

 Applying inversion ($\vec{r}\rightarrow -\vec{r}$, $\vec{k}\rightarrow -\vec{k}$) to reverse molecular handedness, we find  that the displacement and current quadratures in left- (S) and right-handed (R) molecules are connected via $\vec{Q}^{(S)}_{ij}(\vec{k})=\vec{Q}^{(R)}_{ij}(-\vec{k})$ and $\vec{P}^{(S)}_{ij}(\vec{k})=\vec{P}^{(R)}_{ij}(-\vec{k})$.   
The $\vec{Q}(\vec{k})$ quadratures  are related to the lack of rotational symmetry of electron density [around the direction of  $\vec{Q}(\vec{k})$]. In turn, the $\vec{P}(\vec{k})$  quadratures are related to  the lack of rotational asymmetry coming  from the circulating bound state currents  (around the direction of  $\vec{P}(\vec{k})$). The physical origin of the geometric magnetic field are ``twisted'' polarizations and ``curly'' currents. Notably, the currents  induced by the light pulses generate the geometric field, which does not vanish in the molecular frame. $\blacksquare$

The geometric propensity field reflects the geometry of the molecular photoionization dipoles and gives rise to three classes of enantio-sensitive observables, relying on various quadratures of the geometric field  \cite{ordonez_geometric_2021}.  The new enantio-sensitive observables of Class I have been completely overlooked so far. Class I observables can only appear if the current in molecular bound states was excited prior to photoionization. Thus, Class I observables can serve as  messengers of charge-directed reactivity: chemical reactivity driven by ultrafast chiral electron dynamics. The first member of Class I observables is molecular orientation circular dichroism in photoionization  \cite{ordonez_geometric_2021}.
Observables of Class II and III include the PECD (and time-dependent PECD) current and an infinite array of its multipolar versions. Most of these observables have not been studied so far.

\smallskip
\textsf{\noindent\textbf{Concept 5}. \emph{The  geometric field} \cite{ordonez_geometric_2021} is a molecular frame property unique to every molecule.  It underlies vectorial and tensorial enantio-sensitive observables in photoionization, in the electric-dipole approximation. Its flux in photoelectron momentum space quantifies the PEXCD, its integral over all photoelectron momenta quantifies chiral current excited in bound states, its multipole moments characterize multipolar currents, which can be excited by light fields without net spin. The geometric field in photoionization is similar to the Berry curvature in  solids.}
\smallskip

\textbf{Future directions.} The geometric field so far served for us as a heuristic principle for discovering and classifying new enantio-sensitive observables. Future directions can include identification of new members of Classes I-III, establishing the connection  between the topology of the geometric field  and the topology of the respective molecular bound and continuum states, and the 
application of enantio-sensitive molecular orientation  \cite{ordonez_geometric_2021}, which occurs in neural molecules and molecular ions,  for enantio-separation and ultrafast molecular imaging. Other possibilities include  exploiting the analogy between  chiral effects in photoionization and a broad class of topological phenomena in solids, aiming to create  observables which encode both chiral and topological properties of matter  \cite{schwennicke2021enantioselective}, such as the quantized circular dichroism.

\section{The hierarchy of chiral measures}\label{sec:hierarchy}

This section concludes the paper by offering a unified view on chiral measurements and chiral observables as a cornerstone of such measurements, ultimately defining their efficiency.
 
In chiral measurements performed with electromagnetic fields we usually deal with the following objects:  a chiral molecule and either a chiral light (or a combination of electric fields), or a chiral setup. An enantio-sensitive measurement couples the pseudoscalars of the two objects: the chiral molecule and the chiral light and/or electric fields (the 
``chiral reagent'' type of the interaction), or the chiral molecule and the chiral setup (the ``chiral observer'' type of 
the interaction). Remarkably, not only the 
pseudoscalars of all these chiral objects have very similar structure, but they also form a very similar hierarchy associated with the increasing order of non-linear interactions. 

A graphical example of such structural similarity of chiral measures emerges from the comparison of the laser and the setup pseudoscalars. 
Table \ref{tab:hierarchy} shows that in every order of non-linearity the structure of these two pseudoscalars is the same, only 
the specific vectors are different. Namely, the setup pseudoscalar always contains vectors associated with 
the detector axes, which ``substitute'' one or more light vectors of the light pseudoscalar.    

The first row of Table \ref{tab:hierarchy} compares linear chiral measures involving the light and the setup pseudoscalars respectively.  This comparison shows that in PECD the light pseudovector $[\vec{E}^{*}_{\omega}\times\vec{E}_{\omega}]$ substitutes the light pseudovector $\vec{B}^{*}_{\omega}$ in CD, while the detector axis $\hat{z}$ substitutes the light vector $\vec{E}_{\omega}$.

The second-order phenomena involving the light and setup pseudoscalars are shown in the second row of the table.  Here non-linear  CD$^{(2)}$ and  PXCD/ESMW use the same light pseudovector, which is given by the cross product of the electric fields at two different frequencies, but the third vector necessary to form the desired triple product has a very different nature. Indeed, the light vector contributing to the light pseudoscalar in CD$^{(2)}$ is substituted by the detector axis $\hat{z}$ in PXCD/ESMW. The same happens in the fourth order of nonlinearity: the setup pseudoscalar pertinent for tensorial observables such as, e.g., the quadrupole current, substitutes the two laser vectors employed in CD$^{(4)}$ by the detector axes. The presence of the two
detector axes reflects the tensorial nature of the required detection scheme.

Thus, the overall structure of any chiral measure contributing to experimental observables is encoded in pseudoscalar expressions formed by dot and cross products between appropriate vectors. In particular, the chiral measures in the electric  dipole approximation in Table \ref{tab:hierarchy} involve a triple product of three vectors in the lowest order and are subsequently complemented by one scalar product of  two vectors per each subsequent order of non-linearity. 
For example, the two-photon pump-probe photoionization of randomly oriented chiral molecules leads to
the appearance of one additional 
(with respect to PECD) 
scalar product of light fields in the setup pseudoscalar (see the last row of Table \ref{tab:hierarchy}). 

Table \ref{tab:hierarchy} reveals not only the common overall structure of pseudoscalars but also the flexibility in addressing various molecular properties for distinguishing opposite enantiomers. The interchangeability of molecular, light and setup vectors is a great asset for chiral experiments, provided that the vectors are chosen wisely.
Depending on the type of observation, enantio-sensitive response of the same molecular sample can have different strength and different requirements to light pseudovector.

For example, let us compare molecular pseudoscalars for the  linear CD and the non-linear absorption circular dichroism CD$^{(2)}$  \cite{Ayuso2019NatPhot} in Table \ref{tab:hierarchy}. (Note that pseudoscalars of CD$^{(2)}$ also describe the three level enantio-sensitive population transfer  \cite{Eibenberger2017PRL,Perez2017}). We see that in the latter case, instead of relying on the molecular magnetic transition  dipole, one can rely on the cross product of two electric transition  dipoles; instead of relying on the magnetic field one can rely on the cross product of the electric field at two different frequencies. 

One also has freedom in choosing the setup vectors.  They can be constructed not only by introducing detectors for electrons, as done in PECD, but also by using additional electronic or vibrational/rotational degrees of freedom introduced via molecular alignment or coincidence detection involving other electrons,  fragments, etc.

\begin{table*}
\small
\caption{\ Hierarchy of chiral measures.} 
\label{tab:hierarchy}
\begin{tabular*}{\textwidth}{@{\extracolsep{\fill}}llll}
\hline
Phenomenon & Molecular pseudoscalar & Light/setup pseudoscalar\\
\hline
Linear CD  & $[\vec{d}_{f,i}\cdot\vec{m}_{f,i}]$ \cite{Tang2010} & $[\vec{E}_{\omega}^{*}\cdot\vec{B}_{\omega}]$ \cite{Tang2010} \\
PECD  & $\int \mathrm{d}\Omega_k [\vec{k}\cdot(\vec{d}_{\vec{k},i}^{*}\times\vec{d}_{\vec{k},i})]$ \cite{Ordonez2018generalized} &  $[\hat{z}\cdot(\vec{E}^{*}_{\omega}\times\vec{E}_{\omega})]$ \cite{Ordonez2018generalized} \\
    \hline
Non-linear CD$^{(2)}$ & $[\vec{d}_{2,0}\cdot(\vec{d}_{2,1}\times\vec{d}_{1,0})]$ or $\chi^{(2)}\propto \Sigma_{m,n}[\vec{d}_{n,0}\cdot(\vec{d}_{n,m}\times\vec{d}_{m,0})]F_{n,m}$  \cite{Giordmaine1965,fischer_isotropic_2001} & $\{\vec{E}^{*}(\omega_{2,0})\cdot[\vec{E}(\omega_{2,1})\times\vec{E}(\omega_{1,0})]\}$  \cite{Ayuso2019NatPhot}  \\
PXCD/ESMW & $[\vec{d}_{2,0}\cdot(\vec{d}_{2,1}\times\vec{d}_{1,0})]$ \cite{Beaulieu2018PXCD,Ordonez2018generalized}  &  $[\hat{z}\cdot(\vec{E}^{*}_{\omega}\times\vec{E}_{\omega})]$ \cite{,Ordonez2018generalized} \\
\hline
Non-linear CD$^{(4)}$ &
$(\vec{d}_{0,1}\cdot\vec{d}_{1,2})[\vec{d}_{0,2}\cdot(\vec{d}_{2,4}\times\vec{d}_{4,3})]$ or $\chi^{(4)}$ 
&$  (\vec{E}_{\omega}\cdot\vec{E}_{\omega})[\vec{E}^{*}_{2\omega} \cdot (\vec{E}^{*}_{\omega}\times\vec{E}_{\omega})]$ \cite{Ayuso2019NatPhot}\\
Quadrup. currents & $\int \mathrm{d}\Omega_k \{(\hat{k}\cdot \vec{d}_{1,0})[\hat{k}\cdot(\vec{d}_{\vec{k},0}\times\vec{d}_{\vec{k},1})] + (\hat{k}\cdot\vec{d}_{\vec{k},1})[\hat{k}\cdot(\vec{d}_{\vec{k},0}\times\vec{d}_{1,0})] \}$ \cite{ordonez_disentangling_2020} & 
    $(\vec{E}_{\omega}\cdot\vec{E}_{\omega})[\vec{E}_{2\omega}^{*}\cdot(\hat{x}\times\hat{y})]$ \cite{ordonez_disentangling_2020}\\
Permanent quadrup. & $[(Q_{2,2}\vec{d}_{2,1})\cdot (\vec{d}_{1,0}\times\vec{d}_{2,0})] + [(Q_{2,2}\vec{d}_{1,0})\cdot(\vec{d}_{2,1}\times\vec{d}_{2,0})]$ \cite{ordonez_inducing_2021} & $(\vec{E}_{\omega}\cdot\hat{x})[\vec{E}_{2\omega}^{*}\cdot(\vec{E}_{\omega}\times\hat{y})]$ \cite{ordonez_inducing_2021}\\
\hline
PECD$^{(2)}$ (circ. & & $[\vec{E}(\omega_{1,0})\cdot\vec{E}^{*}(\omega_{k,1})]$ \\ pump, circ. probe) & $\int\mathrm{d}\Omega_k (\vec{d}_{1,0}\cdot\vec{d}_{\vec{k},1}^{*})[\vec{k}\cdot(\vec{d}_{1,0}\times\vec{d}_{\vec{k},1})]$ \cite{Ordonez2018generalized}  & $\{\hat{z}\cdot[\vec{E}^{*}(\omega_{1,0})\times\vec{E}(\omega_{k,1})]\}$ \cite{Ordonez2018generalized} \\
TD-PECD (linear & & $[\vec{E}^{*}(\omega_{2,0})\cdot\vec{E}(\omega_{1,0})]$ \\ pump, circ. probe) 
& $\int\mathrm{d}\Omega_k (\vec{d}_{2,0}\cdot\vec{d}_{1,0})[\vec{k}\cdot(\vec{d}_{\vec{k},2}^{*}\times\vec{d}_{\vec{k},1})]$ \cite{Ordonez2018generalized} 
& $\{\hat{z}\cdot[\vec{E}^{*}(\omega_{k,2})\times\vec{E}(\omega_{k,1})]\}$ \cite{Ordonez2018generalized} \\
PEXCD (circ. & & $[\vec{E}^{*}(\omega_{k,2})\cdot\vec{E}(\omega_{k,1})]$ \\ pump, linear probe) 
& $\int\mathrm{d}\Omega_k (\vec{d}_{\vec{k},2}^{*}\cdot\vec{d}_{\vec{k},1})[\vec{k}\cdot(\vec{d}_{2,0}\times\vec{d}_{1,0})]$ \cite{Ordonez2018generalized} 
& $\{\hat{z}\cdot[\vec{E}^{*}(\omega_{2,0})\times\vec{E}(\omega_{1,0})]\}$ \cite{Ordonez2018generalized}
\\
\hline
\end{tabular*}
\end{table*}

Formally,  different dot and cross products in light/molecular/setup pseudoscalars appear as a result of the orientation averaging procedure. Indeed, the isotropy of the molecular sample is the reason why these expressions do not depend on the relative orientations between molecular and setup vectors (e.g. through the terms of the form $\vec{d}_{f,i}\cdot\vec{E}_{\omega}$), but instead only on the relative orientations between either molecular vectors among themselves (e.g. $\vec{d}_{f,i}\cdot\vec{m}_{f,i}$), or 
the setup vectors among themselves (e.g. $\vec{E}_{\omega}\cdot\vec{B}^*_{\omega}$). 

In general, if e.g. the process involves several photons, more vectors become available and a simple \textbf{Rule 1} requires generalization.

\smallskip
\textbf{Rule 7}
\textsf{enantio-sensitive observables pertinent to randomly oriented molecular ensemble  can be written in the general form} \cite{ordonez_molecular_2020, andrews_threedimensional_1977}:
\begin{equation}
v=\sum_{i,j} g_i M_{ij} S_j,
\end{equation}
\textsf{where the $g_i$'s and the $S_j$'s are molecular and setup (or light) pseudoscalars,  respectively, and the $M_{ij}$'s are coupling constants.}
\smallskip

The $g_i$'s result from different possible contractions of Levi-Civita and Kronecker delta tensors with molecular tensors, which then lead to the dot and cross products discussed above. The same applies to the $S_j$'s but using the setup (or light) instead of molecular tensors. In the simplest cases there is a single possible contraction so that the sum over $i$ and $j$ reduces to a single term. This is precisely what occurs in the case of \textbf{Rule 1} in Sec. \ref{sec:chiral_observer}: the measured click, a scalar $v$, corresponding to the projection of the respective vectorial observable $\textbf{v}_{if}$ onto the detector axis $\hat{z}$, converts the laser pseudovector $\textbf{L}$ into the setup pseudoscalar $S$ by projecting $\textbf{L}$ on the detector axis: $S=\textbf{L}\hat{z}$. The resulting expression $v=\textbf{v}_{if}\hat{z}=gS$ is indeed the simplest case of \textbf{Rule 7}.    

\section{Conclusions}
Our perspective on ultrafast chirality presented here offers a unifying framework for understanding and quantifying enantio-sensitive phenomena in interaction of chiral molecules with electromagnetic fields.  This framework is applicable to all  frequency regimes, from microwaves to infrared, to visible, to X-rays and underlies  spectroscopic tools probing electronic, vibronic or rotational states of chiral molecules and detecting photons or photo-electrons. We have described the new concepts of synthetic and locally chiral light, polarization of chirality, and chiral geometric field. These concepts lead to new applications in ultrafast optic, molecular photoionization and nano-photonics. The perspective on synthetic chiral light in confined environments and its applications in nanophotonics will be discussed separately. 

\section*{Acknowledgements}
Discussions with Prof. M. Ivanov, Prof. A. Steinberg, Prof. M. Stockman and Dr. I. Nowitzki were extremely important at different stages of this work and are gratefully acknowledged. A.F.O. and O.S.  gratefully acknowledge the MEDEA Project, which has received funding from the European Union's Horizon 2020 Research and Innovation Programme under the Marie Sk\l{}odowska-Curie Grant Agreement No.
641789. A.F.O. and O.S. gratefully acknowledge support from the DFG SPP 1840 ``Quantum Dynamics in Tailored Intense Fields'' and DFG
Grant No. SM 292/5-2. A.F.O. gratefully acknowledges grants supporting his research at ICFO: Agencia Estatal de Investigación (the R\&D project CEX2019-000910-S, funded by MCIN/ AEI/10.13039/501100011033, Plan Nacional FIDEUA PID2019-106901GB-I00, FPI), Fundació Privada Cellex, Fundació Mir-Puig, Generalitat de Catalunya (AGAUR Grant No. 2017 SGR 1341, CERCA program), and EU Horizon 2020 Marie Skłodowska-Curie grant agreement No 101029393.

\bibliography{references}

\end{document}